\documentclass[twocolumn]{emulateapj}
\usepackage{natbib}
\usepackage{comment}

\def\ba{\begin{eqnarray}}
\def\ea{\end{eqnarray}}
\usepackage{hyperref}
\usepackage{graphicx, amsmath, amsthm, amssymb,color}

\usepackage{ulem}

\shorttitle{Ultra-short-period planets from secular chaos}
\shortauthors{Petrovich, Deibert \& Wu}

\begin{document}

\title{Ultra-short-period planets from secular chaos}

\author{Cristobal Petrovich\altaffilmark{1,2}, Emily Deibert\altaffilmark{3}, \& Yanqin Wu\altaffilmark{3}}

\altaffiltext{1}{Canadian Institute for Theoretical Astrophysics, University of Toronto, 60 St. George Street, Toronto, Ontario, M5S 1A7, Canada}
\altaffiltext{2}{Centre for Planetary Sciences, Department of Physical \& 
Environmental Sciences, University of Toronto at Scarborough, Toronto, 
Ontario M1C 1A4, Canada}
\altaffiltext{3}{Department of Astronomy \& Astrophysics, University of Toronto, 50 St. George Street, Toronto, Ontario, M5S 3H4, Canada}
\begin{abstract}

Over a hundred rocky planets orbiting Sun-like stars in very short orbital periods ($\lesssim1$ day) have been discovered by the {\it Kepler} mission. These planets, known as ultra-short-period (USP) planets, are unlikely to have formed locally, or have attained their current orbits when their birth protoplanetary disks were still present. Instead, they must have migrated in later in life. Here we propose that these planets reach their current orbits by high-eccentricity migration. In  a scaled-down version of the dynamics that may have been experienced by their high mass analog, the hot Jupiters, these planets reach high eccentricities via chaotic secular interactions with their companion planets and then undergo orbital circularization due to dissipation of tides raised on the planet. 
This proposal is motivated by the following observations: 
planetary systems observed by {\it Kepler} often contain several super-Earths with non-negligible eccentricities and inclinations, and possibly extending beyond $\sim$ AU distances;  while only a small fraction of USP planets have known transiting companions, and none closely spaced, we argue that most of them should have companions at periods of $\sim10-50$ days. The outer sibling planets, through secular chaos, can remove angular momentum from the inner most planet, originally at periods of $\sim 5-10$ days. When the latter reaches an eccentricity higher than $0.8$, it is tidally captured by the central star and becomes an USP planet.  
This scenario naturally explains the observation that most USP planets have significantly more distant transiting companions compared to their counterparts at slightly longer periods ($1-3$ days), a feature un-accounted for in other proposed scenarios.
Our model also predicts that USP planets should have: (i) spin-orbit angles,  and inclinations relative to outer planets, in the range of $\sim 10^\circ-50^\circ$; (ii) several outer planetary companions extending to beyond $\sim1$ AU distances, both of which may be tested by {\it TESS} and its follow-up observations.  
\end{abstract}


\section{Introduction} \label{sec:intro}

 Ultra-short-period planets (or USP planets), the rare and enigmatic class of transiting exoplanets with orbital periods shorter than one day, have an unknown origin and are the topic of this study. For reference, a $1$ day orbital period for a solar-type star corresponds to $a \approx 0.02$ AU and a blackbody temperature of $T\approx 2100{\rm K}$. 

About a hundred of these planets have been discovered by the {\it Kepler} transit mission, and they are inferred to exist around $\sim 0.5 \%$ of stars \citep{SO14}, making them slightly less abundant than hot Jupiters 
\citep[Jovian planets orbiting closer than $10$ days, frequency $\sim 1\%$, e.g.,][]{mayor2011,howard2012, Wright12}. 
These planets appear statistically different from the more populous {\it Kepler} systems, their closest analog, in that they are either the only transiting planet in the system, or in cases when they have transiting outer companions \citep[e.g., in Kepler-10 and $\rho$ 55 Cancri systems,][]{Batalha2011,Butler1997,Marcy2002, McArthur2004, Fischer2008}, the latter orbit at periods $\gtrsim 10$ times longer \citep{Steffen2016}, i.e., much further away than in typical {\it Kepler} multi-planet systems.

It is almost certain that USP planets did not form at their current locations.
 These planets fall within the dust sublimation radius for even the most refractive minerals (iron sublimates at $T\approx 2000{\rm K}$). Moreover, the radii of host stars during the pre-main sequence phase were several times larger than their current values \citep[e.g.,][]{Palla1991,DAntona1994} and would have swallowed a number of the closer-in USP planets.

Several formation models have been proposed to explain the origin of these planets, with varying degrees of success. One theory is that they are the exposed cores of giant planets after their gaseous atmospheres have been stripped off by photo-evoporation or tidal forces (\citealt{Jackson2013}, \citealt{Valsecchi2014}, \citealt{Jackson2016}). While there are theoretical objections to this scenario \citep[e.g.,][]{Murray-Clay09}, empirically, \cite{Winn2017} compared the metallicities of  stars harboring USP planets and hot Jupiters and found that they are significantly different, with the hot Jupiters preferentially orbiting around metal-rich stars \citep{Gonzalez1997,Santos2004,Fischer2005}.  Instead, the USP planets' hosts' metallicities are indistinguishable from those of {\it Kepler} planets' hosts, for which there is no notable association with high metallicity (\citealt{Udry2006}, \citealt{Schlaufman2011}, \citealt{Buchhave2012}). They went on to posit that USP planets may be the exposed cores of Neptunes \citep{Valencia2010,Owen2013,Lundkvist2016,Lee2017}. However, even if the latter proposal is correct, one still needs to understand how the planets get so close to the stars in the first place.
Proposals like that in \citet{Mandell2007}, where USP planets are formed from the accretion of material ``shepherded'' inwards by outer giant planets, suffer similar setbacks.

In this work, we propose that most of the ultra-short-period planets were initially the innermost planets in typical multi-planet {\it Kepler} systems. They reach their current orbits from a combination of secular chaotic excitation of their eccentricities and efficient tidal dissipation in the planets at high eccentricities. This is akin to one of the proposals to form hot Jupiters, secular chaos \citep{WL11}. Consider a planetary system with a large number of planets ($N \geq 3$). If these planets are spaced far enough from each other such that their interactions are mostly secular in nature (as opposed to mean-motion resonances), and if the orbits of these planets have some moderate amounts of eccentricity and inclination, secular interactions can become non-periodic and chaotic, leading to diffusive angular momentum transfer among planets that tends to raise the eccentricity and inclination of the innermost planet \citep{laskar96,LW2011,LW2014}.  As this planet's pericenter approaches the central star with an ever-decreasing range, tidal interactions enter at some point. This dissipates the orbital energy of the planet, bringing it to close circular orbits around the star, and snatching it away from the forcing by other planets. Eventually, we are left with a planet that is orbiting at a close range from the star and is dynamically detached from the outer system.

This proposal is motivated by multiple lines of arguments:


\paragraph
{\it 1. analogy with hot Jupiters} there are many observational parallels between USP planets and hot Jupiters. Both are rare classes compared to their more populous cousins: hot Jupiters occur in $\sim 1\%$ of FGK stars, while cold Jupiters occur around $10-15\%$; USP planets occur in $\sim 0.5-1\%$ of stars, while {\it Kepler} systems (planets with radii less than that of Neptune and inward of $400$ days) occur around $30\%$ of stars \citep{Zhu2018}. They also tend to have lower masses compared to these cousins: USP planets have radii $R_p \approx 0.8 - 1.2 R_\oplus$, on the low end of the size spectrum even among those close-in {\it Kepler} planets that have presumably suffered photo-evaporation and are bare rocky cores; while masses of hot Jupiters are a factor of $\sim 2$ or more below the average cold Jupiters \citep[e.g.,][]{DJ2018}.  Lastly, they both lack close neighbors: hot Jupiters have been known to be mostly `lonely' \citep[e.g.,][]{Steffen2012,huang2016}, though lately they have been shown to possess an abundance of distant `friends' \citep{Knutson14}; analogously, most USP planets are either apparently single or have distant (orbits of tens of days) outer companions \citep{Steffen2016}. These similarities propel us to invoke a common mechanism for their formation.

\paragraph
{\it 2. high-multiplicity systems are common} about $30\%$ of stars host low-mass, multi-planet systems inward of 400 days, with an average multiplicity of $3$ \citep{Zhu2018}. 
Further than $400$ days, transit searches are highly incomplete, but there is evidence suggesting that the planet ladder goes on \citep{foreman2016}. Microlensing observations also show that Neptunes are common in long-period orbits, at least around M-dwarfs \citep[e.g.,][]{suzuki2016}.

\paragraph {\it 3. multi-planet systems often have significant eccentricities and inclinations} the {\it Kepler} sample shows that systems hosting three or fewer planets in sub-year orbits (also those most likely to interact secularly) have significant inclination dispersions ($i_{\rm rms}\equiv<i^2>^{1/2}\sim0.1$ ($5.7^\circ$); \citealt{Zhu2018}). These systems are generally observed as single-transiting systems. Interestingly, these very systems also appear to exhibit large eccentricity dispersions ($e_{\rm rms}\equiv<e^2>^{1/2}\gtrsim0.1$; \citealt{xie16}).


This paper is organized as follows. In \S\ref{sec:data} we present the sample of USP planets discovered by the \textit{Kepler} survey and study the possible orbital properties of their outer planetary companions. In \S\ref{sec:sec_chaos} we  set analytical constraints on our proposed mechanism and in \S\ref{sec:setup} we present the results of some numerical experiments. Finally, \S\ref{sec:discussion} provides a discussion of our results and a critique of previous works, and we summarize our main findings in \S\ref{sec:conclusion}.

 \begin{center}
 \begin{figure}
\includegraphics[width=9.5cm]{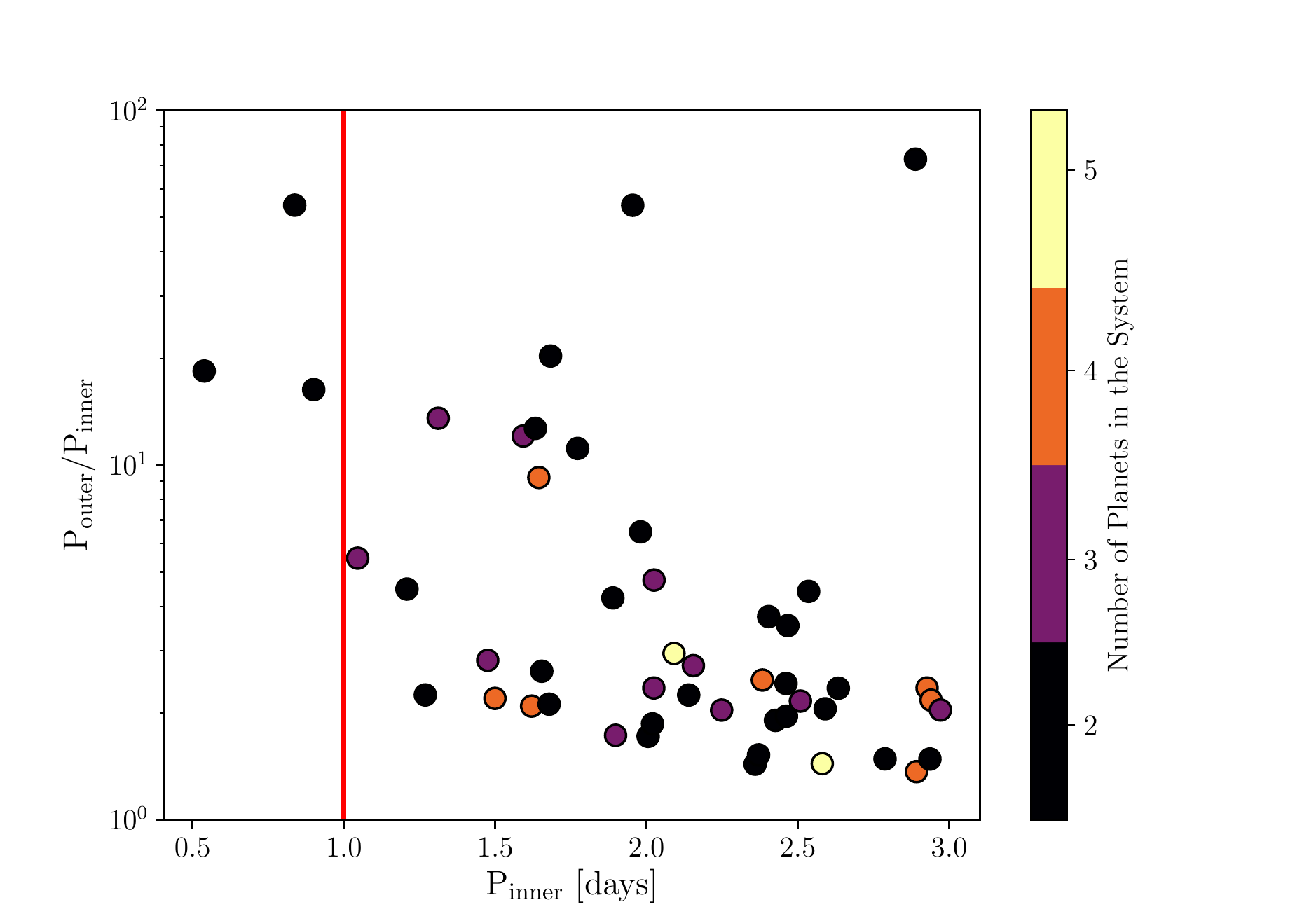}
\caption{Period ratios of adjacent transiting planets where the inner planet has a period less than 3 days and the stellar masses are greater than 0.75 $M_\odot$. The color coding indicates the number of planets transiting in each system. The $3$ (out of 44) ultra-short-period planets (left of vertical line) with outer companions have period ratios $\gtrsim15$, while the planets in slightly longer periods have a broader range, including more compact configurations.}
\label{fig:period_ratio}
\end{figure}
\end{center}

\section{Ultra-short-period planets likely have distant companions} 
\label{sec:neighbors}
\label{sec:data}

Among the currently known USP planets, only a small fraction are in multiple-transiting systems. We use this to infer how likely USP planets are to have distant companions, and the orbital configurations (in terms of orbital periods  and mutual inclinations) of these companions. We conclude that most should have companions, but  the outer companions are either distant ($P\gtrsim20$ days), or are highly mutually inclined ($i_{\rm rms}\gtrsim10^\circ$).  

\begin{table}[htbp!]
\caption{Multiplicity ratio for very-short-period planets with $R_p<4R_\oplus$ discovered by {\it Kepler}}
\centering
\begin{tabular}{c | c | c }
\hline
\hline
inner period & $f_{>1}/f_{1}$= $\#$mult./$\#$sing. & $f_{>1}/f_{1}$ ($M_s\geq0.75M_\odot$)\\
\hline
$<1$ days & $7/55\simeq0.13\pm0.051$& $3/44\simeq0.068\pm0.041$\\
$1-2$ days & $28/53\simeq0.53\pm0.12$ & $18/44\simeq0.41\pm0.11$\\
$2-3$ days & $50/66\simeq0.76\pm0.14$ & $31/54\simeq0.58\pm0.13$\\
\hline
\end{tabular}
\label{tab:results}
\end{table}

\begin{figure*}
\center
\includegraphics[width=17cm]{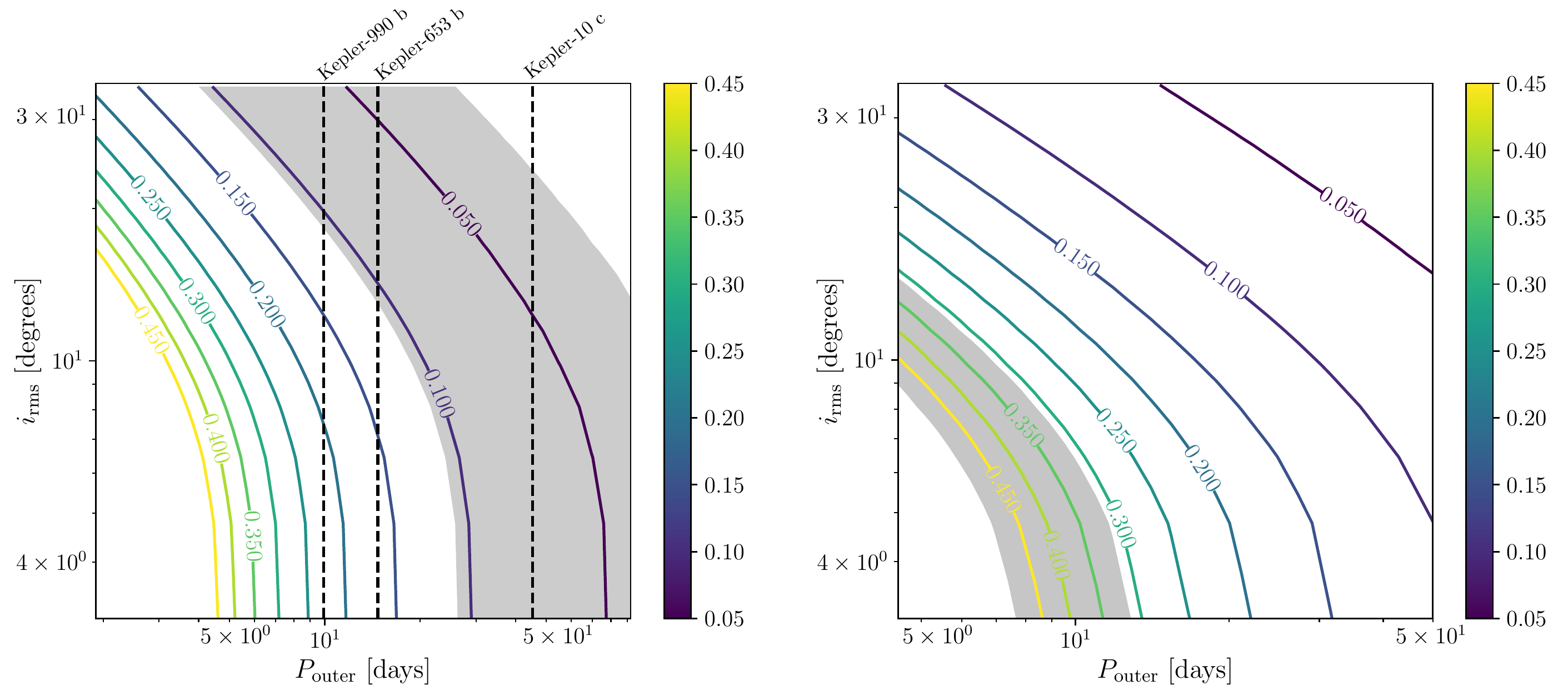}
\caption{Expected  multiplicity ratio ($\left<f_{>1}/f_{1}\right>$ in Equation [\ref{eq:g_sup}]) for different inclination dispersions drawn from a Fischer distribution and periods of outer nearest neighbor. {\it Left panel:} the inner planet is an USP planet ($P_{\rm inner}<1$ day). {\it Right panel:}  we set $P_{\rm inner}\in[1,2]$ days. The inner planets are assumed to have one extra outer companion in the indicated period range.  The shaded region indicates the observed ratio with Poisson error, showing that the outer companions of USP planets are likely {\it beyond} $\sim20$ days.}
\label{fig:g_out_kappa}
\end{figure*}

\subsection{The sample of very-short-period planets ($P<3$ days)}

In Table 1 we show the number of stars with single- and multi-transiting planets discovered by {\it Kepler} with $R_p<4R_\oplus$  and $P\leq3$ days from  the  NASA exoplanet archive as of January 2018\footnote{Confirmed planets from NASA exoplanet archive based on Quarters 1-17,  \url{https://exoplanetarchive.ipac.caltech.edu}}. We notice that the number of stars with single-transiting planets remains roughly constant for the different bins, while the number of stars with multi-transiting planets increases abruptly  with period.  

By looking in more detail at the sample of 7 systems with USP planets in multi-transiting systems, we notice that there seem to be two distinct classes depending on the stellar types: 
\begin{itemize}

\item for $M_s\lesssim0.75M_\odot$ there are 4 systems (out of 11) in preferentially compact configurations, including Kepler-42 (3 planets with $P_{\rm out}/P_{\rm USP}\simeq2.7$), Kepler-32 (5 planets with $P_{\rm out}/P_{\rm USP}\simeq3.9$), and Kepler-80 (6 planets with $P_{\rm out}/P_{\rm USP}\simeq3.1$). The exception to this trend is Kepler-732 (2 planets with $P_{\rm out}/P_{\rm USP}\simeq10$).

\item for $M_s\gtrsim0.75 M_\odot$ there are 3 systems (out of 44) with dynamically detached USP planets, including Kepler-10 (2 planets with $P_{\rm out}/P_{\rm USP}\simeq54$), Kepler-653 (2 planets with $P_{\rm out}/P_{\rm USP}\simeq16$), and Kepler-990 (2 planets with $P_{\rm out}/P_{\rm USP}\simeq18$).  
\end{itemize}

The first class is reminiscent of M-dwarfs which are known to have more miniature systems \citep{Dressing}. Although the USP planets are still more detached compared to other planets, we ignore this class here and focus instead on systems around FGK stars. Thus, we apply an arbitrary cut in host stellar mass of $M_s\geq0.75M_\odot$, which is roughly equivalent to making a cut in effective temperature of $T_{\rm eff}\gtrsim 4600$K. This sample contains 3 USP planets in multi-transiting systems and 44 in single-transiting systems.  

Defining a multiplicity ratio 
\ba 
\frac{f_{>1}}{f_{1}}\equiv \frac{\# \mbox{multi transiting}}{\# \mbox{single transiting}}, \ea
we have $f_{>1}^{\rm USP}/f_{1}^{\rm USP}=3/44\simeq0.068\pm0.041$ for our USP planets. 
In contrast, systems with an inner planet at $P\in [1,3]$ days have $f_{>1}/f_{1}=49/98=0.5\pm0.087$.

We note that the recent study by \citet{adams16} including various transit surveys (Kepler, K2, WASP) finds a similar ratio of $f_{>1}^{\rm USP}/f_{1}^{\rm USP}=11/164\simeq0.067\pm0.021$.  Similarly, the previous work by \citet{SO14} analyzing the \textit{Kepler} data with their own detection pipeline finds a slightly larger number of companions with $f_{>1}^{\rm USP}/f_{1}^{\rm USP}=10/59\simeq0.17\pm0.06$, consistent with our sample without the host star mass cut ($7/55\simeq0.13\pm0.051$) and marginally consistent with the systems with $M_s\geq0.75M_\odot$. 

The multiplicity ratio for USP planets ($0.068$ for our preferred sample) is so low one may worry that many USP planets are truly singles. However, this is unlikely \citep[also see][]{SH15,Steffen2016}.  Even if every USP planet has a companion, but at large periods as is observed for the 3 systems, the chance of observing the companion transit is low. This is made worse if the mutual inclinations are large.  This is quantified in Figure \ref{fig:g_out_kappa} and discussed in detail below. Moreover, since the USP planets are rare with an occurrence rate nearly two orders of magnitude lower than that of average {\it Kepler} systems \citep{SO14}, which are in turn very common and harbor multiple planets \citep{Zhu2018}, it seems reasonable to investigate whether some unusual architecture in the latter systems can lead to USP planet formation, as opposed to assuming that USP planets are a class of their own and are truly single. 

\subsection{Constraints from period ratios}

In Figure \ref{fig:period_ratio} we show the period ratios of adjacent planets for systems with an inner planet inside a $2$ day orbit.  The set of three USP planets have periods ratios  $P_{\rm outer}/P_{\rm inner}\gtrsim15$, while the planets in the period range of $1-2$ days have a median period ratio of $P_{\rm outer}/P_{\rm inner}\sim5$. This result indicates that the USP planets are more dynamically detached than their wider-orbit counterparts.

This result that very-short-period planets have larger periods ratios was previously pointed out by \citet{SF13} and the authors were able to place a boundary at $P_{\rm outer}/P_{\rm inner} \gtrsim 2.3 \left(P_{\rm inner} /\mbox{day}\right)^{-2/3}$. Their results are based on the Quarters 1-12 KOI catalog, while ours are based on the Quarters 1-17 catalog  with confirmed planets and a cut in host star mass (see previous section).

 \subsection{Constraints from multiplicity ratios: $f_{>1}/f_{1}$}

We compute the multiplicity ratio $f_{>1}^{\rm USP}/f_{1}^{\rm USP}$ expected by \textit{Kepler}  assuming that all systems with USP planets have at least one companion inside $\sim50$ days.  

We shall assume that for host stars with detected USP planets, the \textit{Kepler} pipeline has a high detection efficiency at detecting planets with $R\gtrsim1R_\oplus$ in the period range of $P\sim1-50$ days and the ratio  $f_{1}^{\rm USP}/f_{>1}^{\rm USP}$ mainly depends on the transit probabilities.  This is a reasonable assumption as the average detection probability of planets with radii $R_p \gtrsim 2R_\oplus$ ($R_p \gtrsim 1R_\oplus$) with periods $\lesssim 50$ days is near unity ($\gtrsim50\%$) (e.g., \citealt{burke2015,petigura2017}).

We estimate the transit probabilities following the method and notation in \citet{TD11} and provide the necessary details to reproduce our results in the Appendix. From Equation (\ref{eq:g_ups_1}) we compute the expected multiplicity ratio assuming that the USP planet has one outer planetary companion as
\ba
\left<\frac{f_{>1}}{f_{1}}\right>
=\frac{g_{22}(\epsilon_{\rm inner},\epsilon_{\rm outer},\kappa)}{\epsilon_{\rm inner}-
g_{22}(\epsilon_{\rm inner},\epsilon_{\rm outer},\kappa)},
\label{eq:g_sup}
\ea 
where $\epsilon_{\rm inner}=R_s/a_{\rm inner}$,  $\epsilon_{\rm outer}=R_s/a_{\rm outer}$, and $\kappa$ is related to the mean-square value of $\sin{i}$ through Equation (\ref{eqn:kappa}). The function $g_{22}$ is given by Equation (\ref{eq:g22}).
 
In Figure \ref{fig:g_out_kappa} we show the contours of the multiplicity ratio from Equation (\ref{eq:g_sup}) by integrating over the observed range of $\epsilon_{\rm inner}$ for USP planets (left panel) and planets with in the period range of $1-2$ days (right panel). Thus,  $\left<f_{>1}/f_{1}\right>$ depends only on the orbital separation of the outer planets and the inclination dispersion $ \langle \sin^2{i} \rangle $ (or $\kappa$ through Eq. [\ref{eqn:kappa}]).  As expected,  $\left<f_{>1}/f_{1}\right>$ decreases for higher inclination dispersions and longer-period outer planets.
 We include the $1-\sigma$ error bars for the observed ratio $f_{>1}/f_{1}$ from Table 1 with  $M_s\geq0.75M_\odot$ and observe that the USP planets are constrained to large $i_{\rm rms}$ and/or long $P_{\rm outer}$. In particular, if the  $i_{\rm rms}\lesssim10^\circ$, then $P_{\rm outer}\gtrsim25$ days. If  $P_{\rm outer}\lesssim10$ days, then $i_{\rm rms}\gtrsim20^\circ$.

Our estimate of $f_{>1}^{\rm USP}/f_{1}^{\rm USP}$ assumes that there are only two planets in the system. If the intrinsic multiplicity inside $\sim50$ days is higher than 2, then the expected ratio $f_{>1}^{\rm USP}/f_{1}^{\rm USP}$ would increase, demanding for longer periods and higher inclinations to match the observations. Therefore, considering only two planets is a conservative assumption to put constrains on the minimum values of $P_{\rm outer}$.

\begin{figure}[t!]
\center
\includegraphics[width=8.8cm]{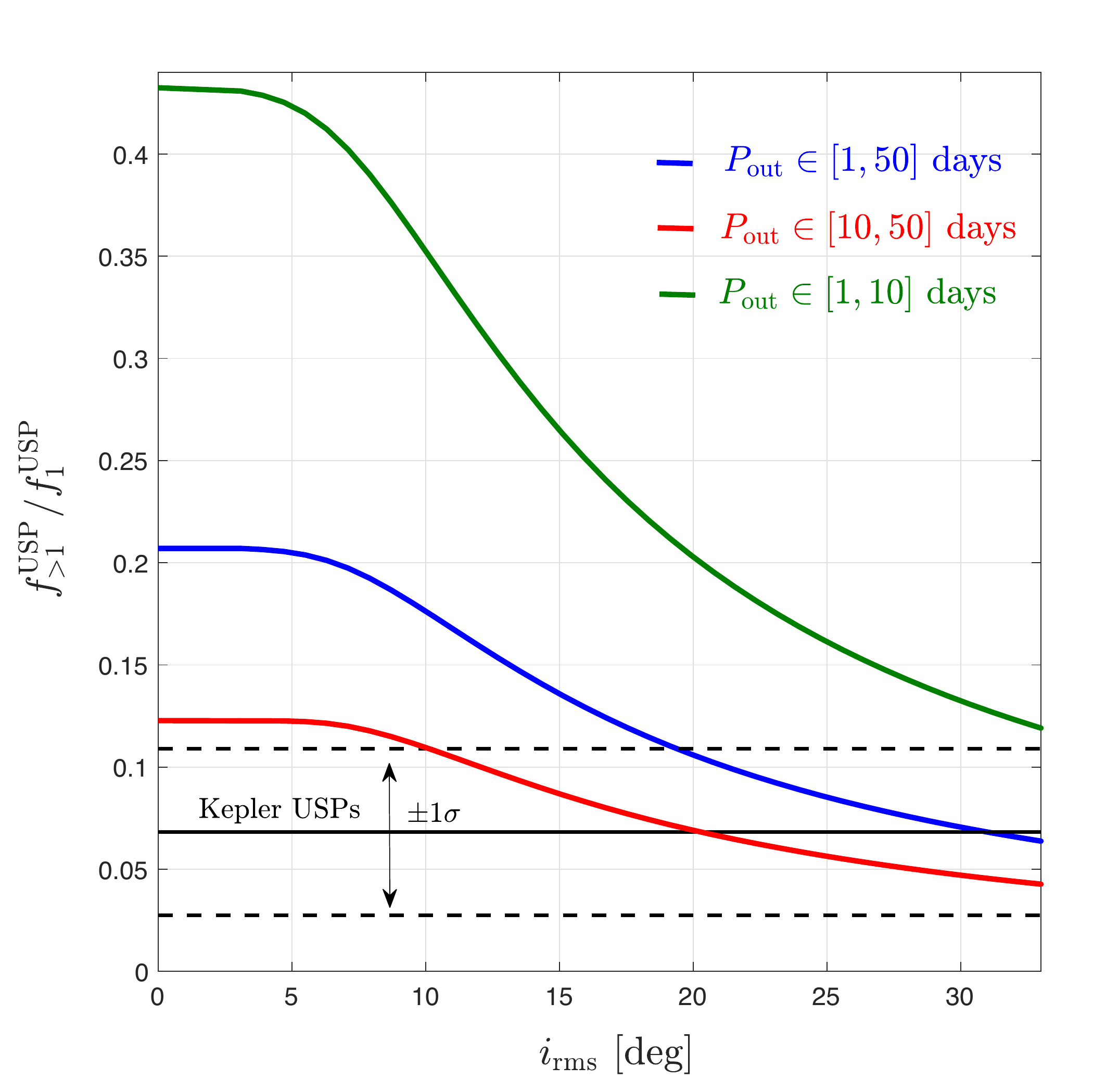}
\caption{
Expected multiplicity ratio $\left<f_{>1}^{\rm USP}/f_{1}^{\rm USP}\right>$, for  different inclination dispersions ($i_{\rm rms}$) drawn from a Fischer distribution assuming that all USP planets have one companion in the period range indicated.  The periods of the outer companion are drawn from the distribution in Equation (\ref{eq:f_epsilon}).}	
\label{fig:i_rms_USP}
\end{figure}

If we further assume that the outer companions do not know about the presence of an inner USP planet and they follow an orbital distribution from the {\it Kepler} sample (e.g., \citealt{TD11})
\ba 
dp(\epsilon_{\rm outer})\propto \frac{(\epsilon_{\rm outer}/0.055)^{0.5}} {1+(\epsilon_{\rm outer}/0.055)^{3.6}}
\frac{d\epsilon_{\rm outer}}{\epsilon_{\rm outer}}, 
\label{eq:f_epsilon}
\ea 
we can integrate $\left<f_{>1}/f_{1}\right>$  in Equation (\ref{eq:g_sup}) over $\epsilon_{\rm out}$. We show the results for these multiplicity ratios in Figure \ref{fig:i_rms_USP} as a function of $i_{\rm rms}$ and for different period ranges.  We observe that if we limit the period range to $P_{\rm outer}<10$ days (green line), then nearly isotropic inclination distributions are required to explain the data. For period-ranges of $1-50$ days (solid blue line), a dispersion of $\sim20^\circ$ or larger is required to explain the data while by limiting ourselves to $10-50$ days, then $\sim20^\circ$ is preferred by the data.

In conclusion, the USP planets have outer companions with typical periods of $\sim20-50$ days and/or $\sim10$ days but which are very highly inclined ($i_{\rm rms}\gtrsim 20^\circ$). If the period of the outer planet is drawn from the observed distribution in {\it Kepler} up to 50 days, then the preferred inclination dispersion is $i_{\rm rms}\gtrsim 20^\circ$.

 \subsection{Comparison with {\it very}-short-period planets ($P\in [1,2]$ days)}

In the right panel of Figure \ref{fig:g_out_kappa} we show the contours of $\left<f_{>1}/f_{1}\right>$ for planets $P\in [1,2]$ days. We observe that, unlike the USP planets (left panel), these systems are consistent with having at least one companion with $P_{\rm outer}\lesssim10$ (even down to $\sim 7$ days) in nearly coplanar orbits ($i_{\rm rms}\lesssim5^\circ$).
 
We note, however, that the planet multiplicity inside 50 days might be higher than 2 planets (see Figure \ref{fig:period_ratio}), in which case planets can be placed at larger orbital distances and still be consistent with the observed  $f_{>1}/f_{1}$. This is unlikely as nearly half of the observed companions are inside 10 days (Figure \ref{fig:period_ratio}).

\section{USP planets produced by secular chaos: analytical preliminaries}
\label{sec:sec_chaos}

Having argued that USP planets likely reside in multi-planet systems, we proceed to discuss other physical constraints in order for secular chaos to produce USP planets. We find that secular chaos can naturally lead to the formation of an USP planet in generic {\it Kepler} systems because:
\begin{itemize}
\item the requisite eccentricity and inclination values are likely common;
\item precession from general relativity and tidal bulges can be overcome for planets initially orbiting beyond $\sim 5$-day obits;
\item secular chaos excites the eccentricities slowly enough that tidal captures can occur, giving rise to planets with final periods of $\sim1$ day.
\end{itemize}
In what follows, we justify each of these statements separately.

\begin{figure}
\center
\includegraphics[width=9cm]{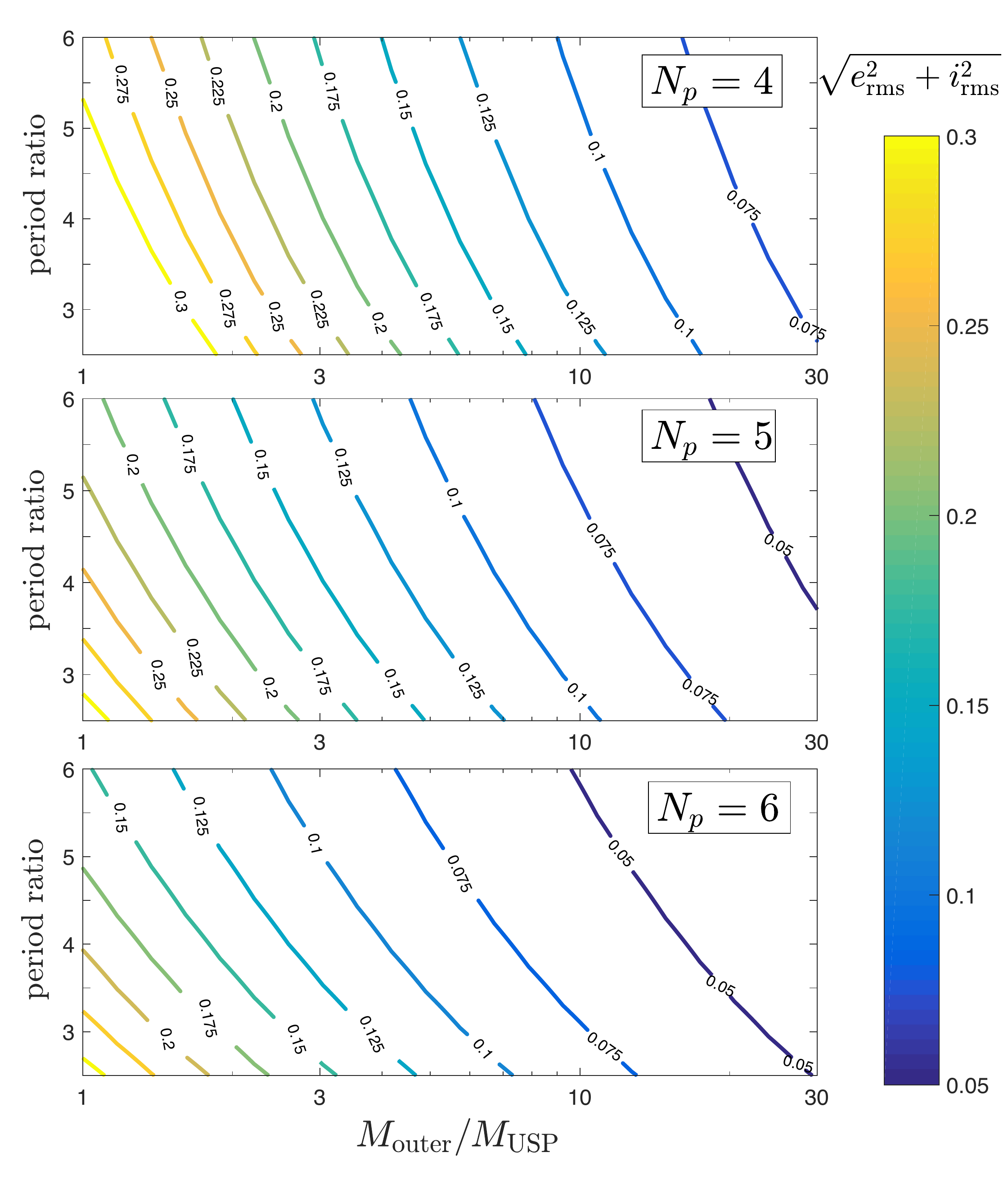}
\caption{Contours of the minimum eccentricity and inclination dispersions expressed as $(e_{\rm rms}^2+i_{\rm rms}^2)^{1/2}$ to produce an USP planet by secular interactions (Eq. [\ref{eq:erms_crit}]) as a function of the mass ratio between the USP planet and the outer planets, and the initial period ratios $P_{\rm i+1}/P_i$ ($i=1...N_{\rm p}$). {\it Upper panel:} total number of planets in the system $N_{\rm p}=4$. {\it Middle panel:} $N_{\rm p}=5$. {\it Lower panel:} $N_{\rm p}=6$. }	
\label{fig:erms}
\end{figure}

\subsection{Required angular momentum deficit to produce USP planets}

The angular momentum deficit (AMD), defined as follows \citep[e.g.,][]{laskar97} 
\ba
\Lambda&=&(GM_s)^{1/2}\sum_{i=1}^{N_{\rm p}} M_i\sqrt{a_i}\left[ 1-\sqrt{1-e_i^2}\cos{i_i}\right]\nonumber\\
&\simeq&(GM_s)^{1/2}\sum_{i=1}^{N_{\rm p}} \onehalf M_i\sqrt{a_i}\left[ e_i^2+i_i^2\right], 
\label{eq:AMD}
\ea 
where $M_s$ is the stellar mass, and $M_i, a_i, e_i, i_i$ the mass, semi-major axis, eccentricity and inclination for planet $i$, describes the deficit in orbital angular momentum relative to that of a coplanar and circular system.  AMD is an important index for the strength of secular interactions. Only when it exceeds a certain threshold can secular chaos occur. Moreover, there must be a minimum amount of AMD for the inner planet to be excited to a highly eccentric orbit and be tidally captured into a tight orbit. We consider this latter constraint below.

Since secular interactions do not modify the orbital energies, AMD is conserved. Thus, for the innermost planet to migrate from $a_1$ to a final circular orbit $a_{\rm 1,f}$ (assuming angular momentum conservation during the circulation process, $a_{\rm 1,f}$  $=a_1[1-e_1^2]$), the minimum AMD required is
\ba
\Lambda_{\rm min}=(GM_s)^{1/2}M_i\left[\sqrt{a_1}-\sqrt{a_{1,f}}\right],
\ea
where we keep the term $\sqrt{a_{1,f}}$, ignored in the case of hot Jupiter migration \citep{WL11} ($\sqrt{a_1}$ is not generally much larger than $\sqrt{a_{1,f}}$ in our scenario). By setting $\Lambda_{\rm min}<\Lambda$ and assuming that all planetary orbits have some typical r.m.s. eccentricity $e_{\rm rms}=\left<e_i^2\right>^{1/2}$ and inclination $i_{\rm rms}=\left<i_i^2\right>^{1/2}$, we get the following condition for migration:
\ba
e_{\rm rms}^2+i_{\rm rms}^2 \leq \frac{2\left[1-\sqrt{a_{\rm 1, f}/a_{\rm 1}}\right]}
{\sum_{i=1}^{N_{\rm p}} \frac{M_i}{M_1}\sqrt{a_{\rm i}/a_1}} \, .
\ea

Typically in our model, the planets start migration from $a_1\sim0.05-0.1$ AU to $a_{\rm 1,f}\sim0.02$ AU, so $\sqrt{a_1/a_{\rm 1, f}}\sim 2$. By assuming that all the outer planets ($i>1$) have the same mass $M_{\rm i}=M_{\rm outer}$ and follow a simple spacing law with constant period ratio  $\mathcal{P}\equiv P_{i+1}/P_i$, the above condition can be expressed as
\ba
\left(\frac{e_{\rm rms}}{0.1}\right)^2&+&\left(\frac{i_{\rm rms}}{0.1}\right)^2
\simeq
\left(\frac{M_{\rm USP}}{1M_\oplus}\right)
\left(\frac{10 M_\oplus}{M_{\rm outer}}\right)\nonumber\\
&&\times\left[\frac{1}{10}\left(\frac{M_{\rm USP}}{M_{\rm outer}}+
\sum_{i=1}^{\rm N_{\rm p}-1}\mathcal{P}^{i/3}\right)
\right]^{-1}.
\label{eq:erms_crit}
\ea
So to produce a given USP planet, the required eccentricities and inclinations are lower if we assume more outer planets which are widely spaced and have higher masses.

In Figure \ref{fig:erms} we show the minimum  $(e_{\rm rms}^2+i_{\rm rms}^2)^{1/2}$ from Equation (\ref{eq:erms_crit}) for $N_p=\{4,5,6\}$ as a function of the mass ratio $M_{\rm outer}/M_{\rm USP}$ and the period ratio $\mathcal{P}$. We observe that for $M_{\rm outer}/M_{\rm USP}\sim1$, the required eccentricities and inclinations are relatively large with $(e_{\rm rms}^2+i_{\rm rms}^2)^{1/2}\sim0.2-0.3$ for $\mathcal{P}\lesssim5$. In turn, by increasing the mass ratio to $M_{\rm outer}/M_{\rm USP}\sim10$ we get  $(e_{\rm rms}^2+i_{\rm rms}^2)^{1/2}\lesssim0.1$, or  $e_{\rm rms}, i_{\rm rms}\lesssim0.07$ in case of equipartition ($e_{\rm rms}\simeq i_{\rm rms}$). 

For reference, there are a few systems where the USP planet has a mass measurement and there is an outer planet with a mass constraint.  Kepler-10 has $M_{\rm outer}/M_{\rm USP}\sim4$ \citep{dum2014,weiss2016}, while Kepler-407b (likely $\sim1M_\oplus$)  has a likely non-transiting outer giant planet \citep{marcy2014}. Other non-\textit{Kepler} systems include 55 Cancri with $M_{\rm outer}/M_{\rm USP}\sim100$  \citet{nelson14} and CoRoT-7 with $M_{\rm outer}/M_{\rm USP}\sim2$ \citep{queloz2009}. We caution that these systems might not be representative of the whole sample as there is a bias towards detecting the most massive planets from radial velocity measurements.

In conclusion, based on the conservation of angular momentum deficit, the formation of an USP planet by secular chaos  roughly requires eccentricity and inclination dispersions at the level of  $(e_{\rm rms}^2+i_{\rm rms}^2)^{1/2}\sim0.1$ for systems with $N_p\gtrsim4$ super-Earths mass planets (USP planets have $\sim$ Earth masses). For Jupiter-mass planets the required dispersions can be much lower.  We shall confirm this result with numerical experiments in \S\ref{sec:setup}.

\subsection{Secular excitation vs apsidal precession forces}
\label{sec:short_range}

We consider whether the diffusive growth in eccentricity for the inner planet, once secular chaos is initiated, can be stalled by other precessional forces. This can limit the maximum eccentricity the planet can reach and prevent USP planet formation. The shortest timescale at which its pericenter distance $r_{\rm p}=a_1(1-e_1)$ is forced to vary is given by the quadrupole forcing from the closest outer companion (planet 2) and is
\ba
\tau_{\rm peri}&\equiv&\left|\frac{r_{\rm p}}{\dot{r}_{\rm p}}\right|\sim P_1\sqrt{1-e_1^2}
\left(\frac{a_{\rm 2}}{a_1}\right)^3 \left(\frac{M_s}{M_{\rm 2}}\right)\nonumber\\
&\simeq&3.7\times10^3\mbox{ yr } 
\left(\frac{0.1\mbox{AU}}{a_1}\right)^2
\left(\frac{a_{\rm 1,f}}{0.02\mbox{AU}}\right)^{1/2}
\nonumber\\
&&\times\left(\frac{15~M_\oplus}{M_2}\right)\left(\frac{a_2}{0.24\mbox{ AU}}\right)^3,
\label{eq:tau_p}
\ea where $a_1$ is the starting distance for the USP planet and we have evaluated the fiducial values of $M_2$ and $a_2$ as those for Kepler-10c \citep{weiss2016}. If the planet reaches $e_1\sim1$, but its inclination is still moderate ($\lesssim40^\circ$) then the quadrupole forcing vanishes and $\tau_p$ has a longer (octupole) timescale by a factor of $\sim a_2/(e_2 a_1)$.

\subsubsection{Relativistic precession}
\label{sec:gr_quenching}

The relativistic precession can change the argument of periapsis in a characteristic timescale given by
\ba
\tau_{\rm GR}&=&\frac{c^2a_1 (1-e_1^2)}{3 GM_s}P_1\nonumber\\
&\simeq&2\times 10^4\mbox{ yr } \left(\frac{a_1}{0.1\mbox{AU}}\right)^{3/2}
\left(\frac{a_{\rm 1,f}}{0.02\mbox{AU}}\right).
\ea
As the eccentricity increases and $a_{\rm 1,f}=a_1(1-e^2)$ decreases, the above apsidal precession rate rises, leading to a quenching of the secular perturbations from the outer planet when $\tau_{\rm peri}=\tau_{\rm GR}$, which occurs at
\ba
a_{\rm f,GR}\sim 7\times10^{-4}\mbox{AU}
\left(\frac{0.1\mbox{AU}}{a_1}\right)^{7}
\left(\frac{15~M_\oplus}{M_2}\right)\left(\frac{a_2}{0.24\mbox{ AU}}\right)^6.\nonumber\\
\label{eq:a_gr}
\ea
For instance, for Kepler-10b to reach its current semi-major axis $a_{1,f}\simeq 0.0167$ AU, it should have started migration from $a_1\gtrsim0.06$ AU ($P_1\gtrsim5.4$ days), which we determine by setting $0.0167\mbox{ AU}>a_{\rm f,GR}$.

\begin{figure*}[t!]
\center
\includegraphics[width=17cm]{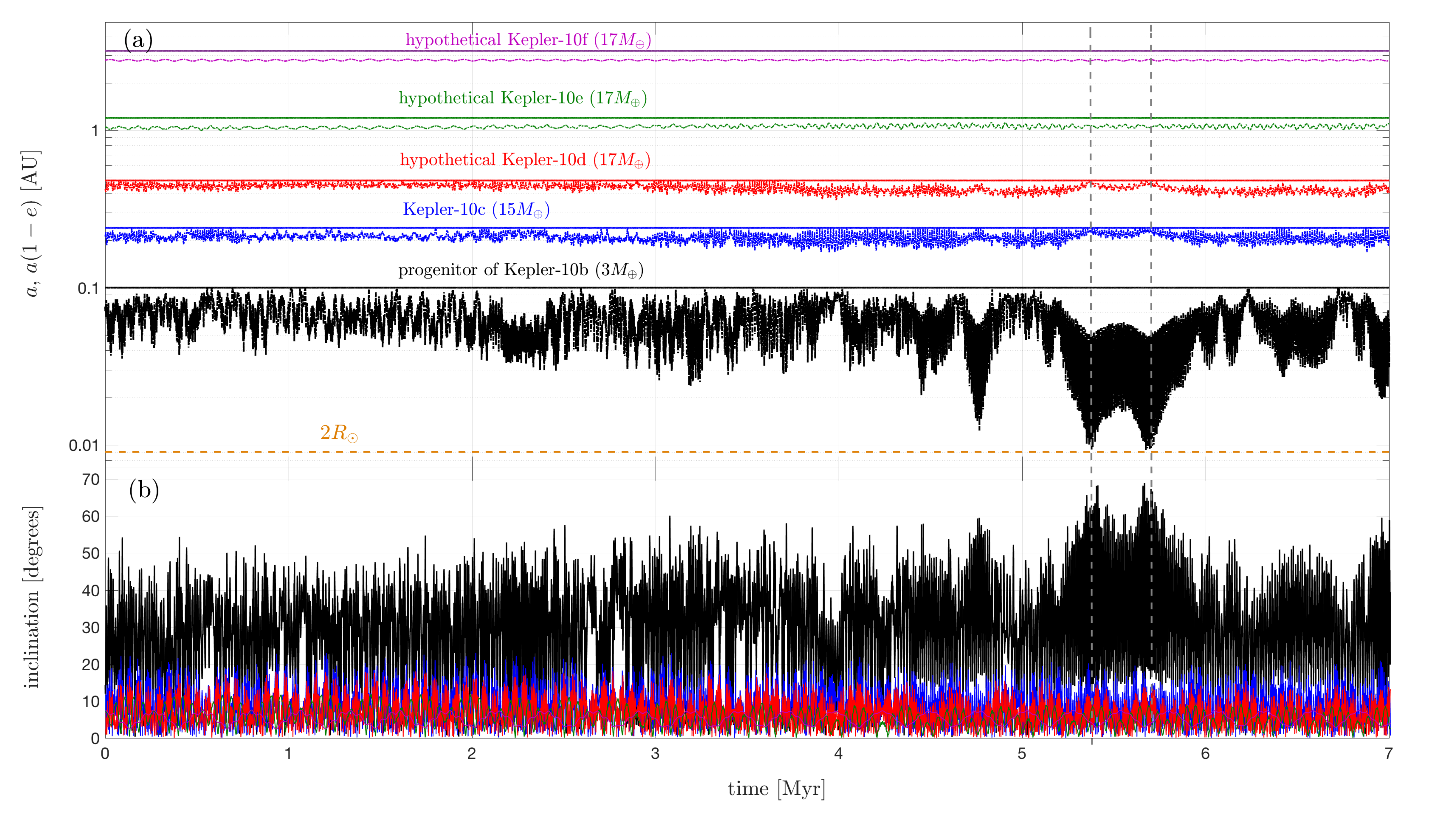}
\caption{Example of a possible evolution to produce the Kepler-10 system. The upper panel (a) depicts planet semi-major axis, and pericenter distance ($a_i[1-e_i]$), while the lower panel that of inclinations.
Due to secular chaos with the companions, planet Kepler-10b reaches $a_1(1-e_1^2)\lesssim0.02$ AU after $\sim 5$ Myr, at which time tidal damping ($f$-mode diffusion or equilibrium tides) can potentially shrink the orbit to a final period $\lesssim1$ day. At the moment that the planet reaches its maximum eccentricity (indicated by gray vertical lines), its inclination is also the highest ($i\sim70^\circ$), while the eccentricities of planets c and d are low ($\lesssim0.05$). The initial eccentricities and inclinations for planets c to f are assumed to be $e=0.12$ and $i=7^\circ$, while higher values ($e_1=0.22$ and $i_1=12^\circ$) are assumed for Kepler-10b to speed up its chaotic diffusion.}
\label{fig:K10_evol}
\end{figure*}


\subsubsection{Precession by tidal bulges}

The tidal deformation of both the planet and the star by their mutual gravitational perturbations leads to apsidal precession. Assuming that $e_1\sim1$, the characteristic timescale for the tidal bulge on the planet is given by \citep{Sterne39}
\ba
\tau_{\rm tide,p}&\simeq&
\frac{16}{315k_{L,p}}P_1
\left(\frac{M_1}{M_s}\right)
\left(\frac{a_{\rm 1,f}}{R_1}\right)^5\nonumber\\
&\simeq&4\times 10^5\mbox{ yr }
\left(\frac{0.3}{k_p}\right)
\left(\frac{a_1}{0.1\mbox{AU}}\right)^{3/2}
\left(\frac{M_1}{1M_\oplus}\right)\nonumber\\
&&\left(\frac{1R_\oplus}{R_1}\right)^5
\left(\frac{0.02\mbox{AU}}{a_{\rm 1,f}}\right)^5,
\ea
where $k_{L,p}$ is the tidal Love number and we scale it by that of the Earth.
Similarly, the tidal bulge on the star gives rise to
\ba
\tau_{\rm tide,s}&\simeq&
\frac{16}{315k_{L,s}}P_1
\left(\frac{M_s}{M_1}\right)
\left(\frac{a_{\rm 1,f}}{R_s}\right)^5\nonumber\\
&\simeq&6\times10^7\mbox{ yr }
\left(\frac{0.014}{k_s}\right)
\left(\frac{a_1}{0.1\mbox{AU}}\right)^{3/2}
\left(\frac{1M_\oplus}{M_1}\right)\nonumber\\
&&\left(\frac{R_s}{1R_\odot}\right)^5
\left(\frac{0.02\mbox{AU}}{a_{\rm 1,f}}\right)^5\, ,
\label{eq:a_tide}
\ea  
where $k_{L,s}$ is the tidal Love number of the star and we scale it by the solar value.
Thus, the precession rate is generally dominated by the tidal bulge on the planet instead of the star. As argued by \citet{liu15}, the maximum eccentricity allowed for the tidal bulges in three-body interactions is reached when $\tau_{\rm tide,p}\sim 0.1\tau_{\rm peri}$, which occurs at 
\ba
 a_{\rm f,tide}&\sim& 4\times10^{-3}\mbox{ AU }
\left(\frac{R_1}{1R_\oplus}\right)^{10/9}
\left(\frac{0.1\mbox{AU}}{a_1}\right)^{7/9}\nonumber\\
&&\left(\frac{a_2}{0.24\mbox{AU}}\right)^{2/3}
\left(\frac{1M_\oplus}{M_1}\cdot\frac{15M_\oplus}{M_2}
\right)^{2/9}.
\ea
This expression sets the minimum $a_1(1-e_1^2)$ allowed by tidal bulges. We note that for Kepler-10b with $R_p\simeq1.46R_\oplus$ and $M_1\simeq3.3M_\oplus$, the current location implies that migration should have started from $a_1 \gtrsim0.045$ AU ($P_1\gtrsim3.5$ days), comparable to that obtained from relativistic precession.

\subsection{Tidal captures}

We assess whether the USP planets can attain their current detached orbits by tidal decay and whether these can prevent their tidal disruptions. We describe the roles of equilibrium tides and dynamical tides separately.

\subsubsection{Equilibrium tides}

Since the planet's spin synchronizes with the orbit in short timescales compared with migration timescales, the planet needs to be in an eccentric orbit for friction to extract orbital energy.

We describe the tidal effects on the orbital evolution of the planet using the weak friction theory of equilibrium tides (e.g., \citealt{hut81}), according to which the rate of decay of the semi-major axis for a pseudo-synchronized planet can be written as
\ba
\left(\tau_{a}^{\rm planet}\right)^{-1}&\equiv&\left|\frac{\dot{a_1}}{a_1}\right|
=6k_{L,p}\tau_p \left(\frac{GM_s}{a_{\rm 1,f}^3}\right)
\left(\frac{R_1}{a_{\rm f}}\right)^5\nonumber\\
&&\times\sqrt{\frac{a_{\rm 1,f}}{a_1}}
\frac{M_s}{M_1}\mathcal{F}(e_1),
\nonumber\\
\label{eq:tau_a_hut}
\ea
where $a_{\rm 1,f}=a_1(1-e_1^2)$ is the final circularization radius, $k_{L,p}$ is the tidal Love number, $\tau_p$ is the tidal lag time (assumed constant in the weak friction theory), and 
\ba
\label{eq:F_e}
\mathcal{F}(e)&=&1+\frac{31}{2}e^2+\frac{255}{8}e^4+\frac{185}{16}e^6+\frac{25}{64}e^8\nonumber\\
&&-\frac{\left(1+\frac{15}{2}e^2+\frac{45}{8}e^4+\frac{5}{16}e^6\right)^2}{1+3e^2+\frac{3}{8}e^4}.
\ea

From Equation (\ref{eq:tau_a_hut}) the timescale to form an USP planet with final period $P_{\rm 1, f}=1$ days ($a_{\rm 1,f}=a_1[1-e_1^2]\simeq$0.02 AU)  starting from a period $P_1$ around a Sun-like star becomes
\ba
\tau_{a}^{\rm planet}&\simeq&1.37\times10^3~\mbox{yr}~
\frac{1}{k_{L,p}\tau n_{\rm1, f}}
\left(\frac{1R_\oplus}{R_1}\right)^5
\left(\frac{M_1}{1M_\oplus}\right)\nonumber\\
&&\times\left(\frac{P_{\rm 1,f}}{1\; \mbox{day}}\right)^{4}
\left(\frac{P_1}{10\;\mbox{day}}\right)^{1/3}
\left(\frac{\mathcal{F}(0.9)}{\mathcal{F}(e_1)}\right)\nonumber\\
&\simeq&10^5~\mbox{yr}~\left(\frac{0.3}{k_{L,p}}\right)
\left(\frac{600~\mbox{s}}{\tau_p}\right)
\left(\frac{1R_\oplus}{R_1}\right)^5
\left(\frac{M_1}{1M_\oplus}\right)\nonumber\\
&&\times\left(\frac{P_{\rm 1,f}}{1\; \mbox{day}}\right)^{5}
\left(\frac{P_1}{10\;\mbox{day}}\right)^{1/3}
\left(\frac{\mathcal{F}(0.9)}{\mathcal{F}(e_1)}\right).
\ea
For Earth-like planets we assume $\tau_p \sim600$ s \citep{lambeck77,neron97}.

Using the tidal quality factor $Q_p\equiv(\tau_p \omega)^{-1}$ with $\omega=n_{\rm 1,f}\equiv(GM_s/a_{\rm 1,f}^3)^{1/2}$ we obtain
\ba
\tau_{a}^{\rm planet}&\simeq&1.2\times10^5~\mbox{yr}~
\left(\frac{Q_p/k_{L,p}}{100}\right)
\left(\frac{1R_\oplus}{R_1}\right)^5
\left(\frac{M_1}{1M_\oplus}\right)\nonumber\\
&&\times\left(\frac{P_{\rm 1,f}}{1\; \mbox{day}}\right)^{4}
\left(\frac{P_1}{10\;\mbox{day}}\right)^{1/3}
\left(\frac{\mathcal{F}(0.9)}{\mathcal{F}(e_1)}\right).
\label{eq:tau_a_Q}
\ea
Evidently this timescale is short enough so that if planets can attain eccentricities large enough to reach $P_{\rm 1,f}=P_1(1-e_1^2)^{2/3}\lesssim1$ day ($a_{\rm 1mf}\lesssim0.02$ AU), then circularization is possible. However, since relativistic precession and tidal bulges do not efficiently limit the eccentricity growth, the pericenter can continue shrinking until the planet gets tidally disrupted.

The disruption can be prevented by a tidal capture, meaning that the planet can shrink its orbit significantly by tidal dissipation before the pericenter continues to approach the disruption distance. This possibility seems promising because, as shown by \citet{munoz2016}, rocky planets can survive secular migration for a wide range of parameters compared to gaseous planets.

By equating $\tau_{\rm peri}$ (Eq. [\ref{eq:tau_p}]) and  $\tau_{a}^{\rm planet}$ (Eq. [\ref{eq:tau_a_Q}]) we get that a tidal capture occurs at
\ba
a_{\rm f, capture}&\simeq&0.011\mbox{ AU } 
\left(\frac{R_1}{1R_\oplus}\right)^{10/11}
\left(\frac{a_2}{0.24\mbox{ AU}}\right)^{6/11}\nonumber\\
&&
\left(\frac{0.1\mbox{AU}}{a_1}\right)^{5/11}
\left(\frac{100}{Q_p/k_{L,p}}\cdot\frac{1M_\oplus}{M_1}\cdot\frac{15M_\oplus}{M_2}
\right)^{2/11}.
\label{eq:a_capture}
\ea

 For Kepler-10b with $R_1=1.47R_\oplus$ we get the following condition
 for a capture at its current location (i.e., $a_{\rm f, capture}=0.0167$ AU)
 \ba
 \left(\frac{a_1}{0.1\mbox{AU}}\right)^{5/11}\left(\frac{Q_p/k_{L,p}}{100}
\right)^{2/11}\simeq0.56.
 \ea
Thus, it is possible that Kepler-10b has achieved its current location
by a tidal capture if it started migration from $a_1\sim0.06$ AU.

We stress that the final semi-major achieved by a tidal capture is the {\it minimum } value allowed by tidal dissipation in the planet. The actual value for a planet undergoing high-eccentricity migration might be longer for the following separate reasons:
\begin{enumerate}
\item tidal dissipation shrinks the orbit after several secular cycles, not one as it is assumed in a tidal capture;
\item the secular forcing can have a longer timescale than the one used above ($\tau_{\rm peri}$ in Eq. [\ref{eq:tau_p}]) because it can be driven by the octupole moment, not the quadrupole, from the outer planetary orbit. If so, $\tau_{\rm peri}$ increases by a factor of $\sim a_2/(e_2a_1)$, increasing $a_{\rm f, capture}$ by a factor of $\sim [a_2/(e_2a_1)]^{2/11}$.
\end{enumerate}

\subsubsection{Dynamical tides: diffusive $f$-mode excitation preventing disruptions}

As discussed above, the tidal dissipation rate in the planet from equilibrium tides might be efficient enough to tidally capture the proto-USP planet and prevent its disruption. However, the dissipative properties of the short-period planets are quite uncertain,  
and the values of $Q_p/k_{L,p}$ can potentially be large enough that they invalidate our previous statement \citep[e.g., GJ 876-d, ][]{{PB2018}}. 

 Fortunately, even in the limit of an inviscid planet ($Q_p \rightarrow \infty$), there is salvation. It was recently pointed out by \citet{VL2018} and \citet{Wu2018} that, regardless of the planetary dissipative properties, a planet in a highly eccentric orbit can diffusively excite its spherical-degree 2 fundamental mode (f-mode) to near-unity amplitudes. When this happens, that nonlinear effects can set in and effectively convert mode energy to heat. As a result, the orbital energy is lost and the orbit shrinks in a short timescale. This circumvents the difficulty surrounding the uncertain dissipation of the equilibrium tide.

As shown by \citet{Wu2018}, the mode excitation can enter the diffusive regime and the mode energy can grow linearly in time when the pericenter distance $r_p=a_1(1-e_1)$ reaches
\ba
r_p\lesssim4 R_p\left(\frac{M_s}{M_1}\right)^{1/3}\simeq0.012\mbox{ AU}
\left(\frac{5~ {\rm g/cm^3}}{\rho}\right)^{1/3},
\label{eq:rpdyn}
\ea
with some small corrections that depends on the $f$-mode period. Here $\rho$ is the density of the planet. At this point, the orbit shrinks with a rapid timescale of (Eq. [17] in \citealt{Wu2018})
\ba
\tau_{a,{\rm f\mbox{-}mode}}\sim 10^2 \mbox{ yr}\left(\frac{a_1}{0.1\mbox{AU}}\right)^{3}.
\ea
Equation (\ref{eq:rpdyn}) also yields the final orbital semi-major axis by $a_{1,f} \approx 2 r_p$. 
This is because $r_p$ also constitutes an impassable wall for the planet migration. The timescale for mode growth drops steeply as $r_p^{-21}$, any further decrease of the pericenter distance beyond that in Equation (\ref{eq:rpdyn}) brings exponentially faster decay in orbital energy. This efficiently decouples the planet from its secular perturber. When the planet's orbit circularizes from $e_1 \approx 1$, we obtain $a_{1,f} \approx 2 r_p$, or
\ba
a_{\rm f,f-mode}\lesssim8 R_p\left(\frac{M_s}{M_1}\right)^{1/3}
\simeq0.024\mbox{ AU}
\left(\frac{5~ {\rm g/cm^3}}{\rho_p}\right)^{1/3},
\label{eq:a_fmode}
\ea
which corresponds to a final period $\lesssim1.3$ days. 

In conclusion, the diffusive excitation of the $f$-mode allows for rapid orbital migration when the planet reaches inwards of the value in Equation (\ref{eq:rpdyn}). This prevents any secularly migrating planets from being pushed even closer to the star and suffering  the fate of tidal disruption. We end up with an USP planet that lies around the observed distances, subject to uncertainties in f-mode period, planet density, etc. Subsequent tidal dissipation, possibly via equilibrium tides (likely for solid planets), may eventually circularize the orbit.

\begin{figure*}[t!]
\center
\includegraphics[width=16cm]{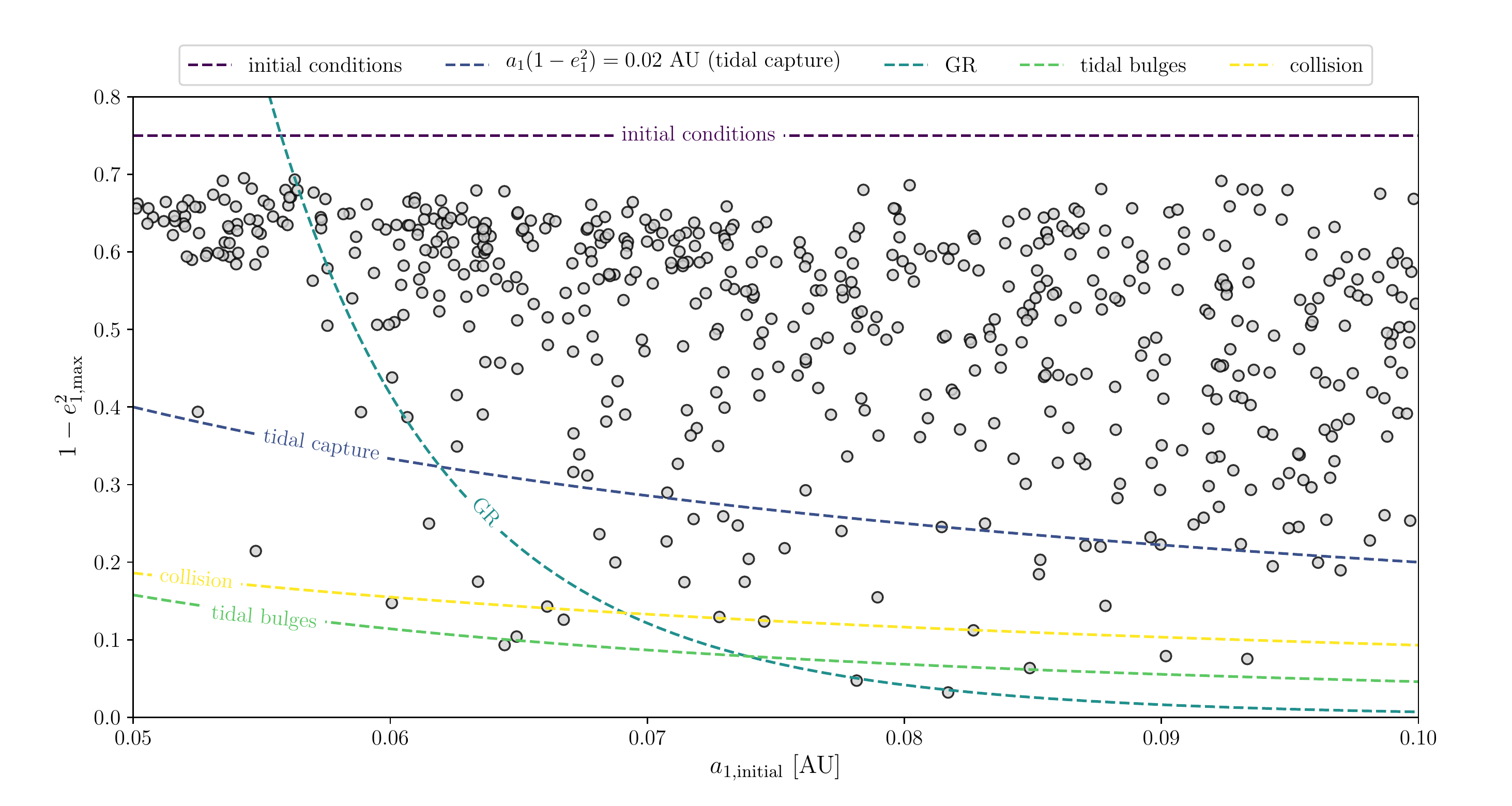}
\caption{Maximum eccentricities reached by a Kepler-10b-like planet after $10^6$ yr for different initial semi-major axes. We place planet b with an initial high eccentricity  ($e_1=0.5$ , purple horizontal line at 0.75) and inclination ($i_1=28.6^\circ$) to artificially speed up the diffusion to large $e_1$. The outer planets have eccentricities and inclinations drawn from a Rayleigh distribution with $\sigma_e=\sigma_i=0.1$ (rms values of 0.14) and have the same spacing and masses as in Figure \ref{fig:K10_evol}. The remaining orbital elements ($f$, $\omega$, $\Omega$) are drawn randomly. The integrations include relativistic precession and non-dissipative tidal bulges, and we indicate the maximum eccentricities allowed by relativistic precession (Eq. [\ref{eq:a_gr}]) and tidal bulges (Eq. [\ref{eq:a_tide}]). For reference we indicate the threshold to become an USP planet of $a_1(1-e_1^2)<0.02$ AU. The integration timestep is $0.1$ day.}	
\label{fig:pop_emax}
\end{figure*}

\section{Numerical experiments} 
\label{sec:setup}

We explore numerically the validity of our analytical estimates in the previous section by running direct numerical integrations of a possible initial configuration of the Kepler-10 system. Our calculations should be taken as a proof of concept and not as a detailed population synthesis study, which is beyond the scope of this work.

\subsection{Code}

All integrations were performed using the {\sc \tt WHFAST} integrator \citep{RT15} in the open-source {\sc \tt REBOUND} N-body package \citep{Rein2011}. We include the effects from relativistic precession and apsidal  precession from tidal bulges from the  {\sc \tt REBOUNDx}\footnote{\url{https://github.com/dtamayo/reboundx}} library with the option {\it gr-potential} and Love numbers $k_{L,p}=0.3$ and $k_{L,s}=0.014$ for the planets and the star, respectively  (Tamayo et al., in prep.).

 Our experiments do not include tidal dissipation and we use the maximum eccentricity $e_{\rm1, max}$ as a proxy for the potential formation of an USP planet: planets reaching $a_1(1-e_{\rm1, max}^2)\lesssim0.02$
 AU can be tidally captured to a final semi-major axis of $\lesssim0.02$ AU ($P_1\lesssim1$ day).
 
\subsection{Example and orbital architecture}

We shall assume that the Kepler-10 system has planets beyond $\sim100$ days and that these planets have Neptune masses, similar to the mass of Kepler-10c \citep{weiss2016}. We place these hypothetical Kepler-10x planets with periods $\simeq122$ days, $\simeq480$ days, and $\simeq2100$ days. Although our choice of orbital configurations is arbitrary, its general architecture is broadly consistent with the bulk  of planetary systems in the \textit{Kepler} sample:
\begin{itemize}
\item there are three planets inside $\sim400$ days, consistent with the average number of planets in this range\footnote{Planet Kepler-10c has TTVs, indicating the presence of a third and unseen planet in the  system  \citep{weiss2016}.} \citep{Zhu2018};
\item the Neptune-size planets in the range of $\sim2-25$ years are at least as common than their sub-year period counterparts \citep{foreman2016}.
\end{itemize}

In Figure \ref{fig:K10_evol}, we show the evolution of one possible progenitor of the Kepler-10 system.  Here, Kepler-10b starts at $a_1=0.1$ AU and its eccentricity and inclination evolve to large values due to secular chaotic diffusion driven by the outer planets \citep[e.g.,][]{laskar96,WL11}.  As expected from these secular perturbations, the semi-major axes of all the planets remain constant (indicated by the horizontal solid lines).

The planet b reaches a maximum eccentricity of $e_1\simeq0.9$ after $\sim5$ Myr, so its pericenter distance becomes $a_1(1-e_1)\simeq2R_\odot$ and it could be tidally captured to become an USP planet ($a_1[1-e_1^2]\lesssim0.02$ AU).  At this point (indicated by vertical dashed lines) its inclination is also largest ($i_1\sim70^\circ$) so if the planet were tidally captured, it would likely have a large inclination relative to the outer planets and the host star spin axis. 

In our picture, secular chaos is a means to stabilize the system by reducing its overall angular momentum deficit \citep{WL11}. If planet Kepler-10b  were tidally captured at  $e_{\rm 1,max}$, then the orbits of the outer planets would gain angular momentum and become more circular. In fact, at $e_{\rm 1,max}$ (vertical dashed lines) the planets c and d have eccentricities of $e\simeq0.05$ compared to their averages of $\sim0.15$ during the rest of the evolution.

\subsection{Maximum eccentricities allowed by short-range forces}

In Figure \ref{fig:pop_emax} we show the maximum eccentricities reached for 500 integrations with the same architecture as Figure \ref{fig:K10_evol} but changing the initial semi-major axis of planet b to illustrate the effect from short-range forces and compare with our analytical estimates from \S\ref{sec:short_range}.

We initialize  the orbit of the innermost planet with $e_1=i_1=0.5$ to artificially speed up the diffusion to large eccentricities as we use a short integration timestep (0.1 day) and relatively small maximum integration time (1 Myr). Our goal is to properly resolve very large eccentricities ($e_1\gtrsim0.9$). The subsequent experiments do not assume initial large $e,i$ for the innermost planet.

We observe that the distribution of $1-e_{\rm 1,max}^2$ looks roughly uniform for systems with initial semi-major axes $\gtrsim0.075$ AU, while it significantly shrinks towards the initial values ($1-e_1^2=0.75$) for $\lesssim0.06$ AU as expected from the short-range forces.

From \S\ref{sec:short_range}  (Eqs. [\ref{eq:a_gr}]) and (Eq. [\ref{eq:a_tide}]) we expect that tidal bulges limit the maximum eccentricity for $a_1\gtrsim 0.07$ AU, and we observe that  apart from a few exceptions, the planets are indeed above the line from the tidal bulges (green lines). For $a_1\lesssim 0.07$ AU, relativistic precession is expected to dominate and the bulk of the experiments are roughly above this boundary.

Finally, we notice that both relativistic precession and tidal bulges allow for the formation of USP planets for $a_1\gtrsim0.06$ AU: the line at $a_1(1-e_1^2)=0.02$ AU is above GR and tides in this semi-major axis range.

\begin{figure}[h!]
\center
\includegraphics[width=8.8cm]{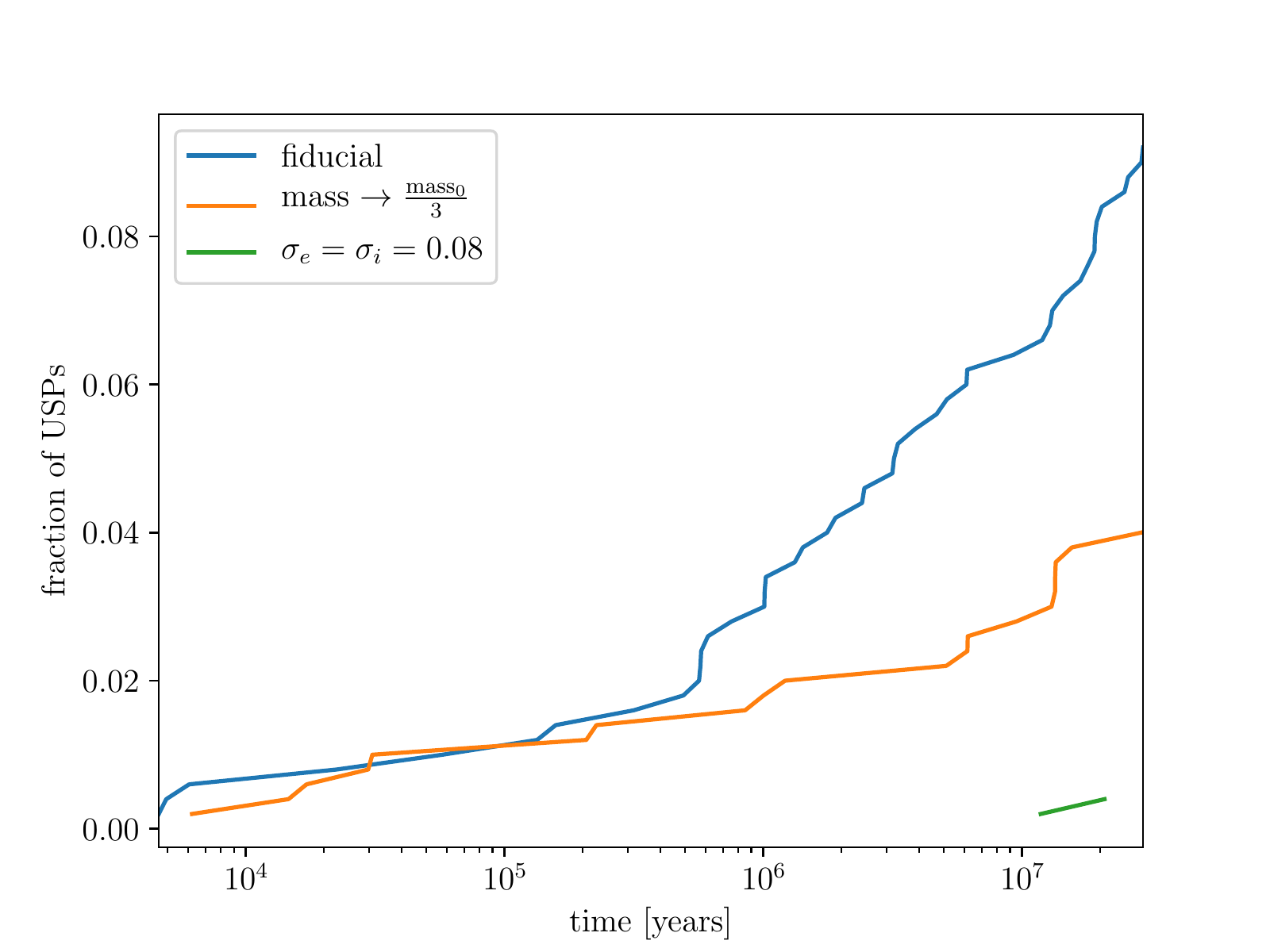}
\caption{Fraction of USP planets formed (i.e., reaching $a_1(1-e_1^2)<0.02$ AU) as a function of time for Kepler-10 systems where the innermost planet starts with a random semi-major axis in $a_1\in[0.06,0.1]$ AU.
In our fiducial run (blue line), all planets have initial eccentricities and inclinations drawn from a Rayleigh distribution with parameter $\sigma=0.1$ ($e_{\rm rms}=i_{\rm rms}=0.14$) and the masses and semi-major axes of the outer planets are the same as in Figure \ref{fig:K10_evol}. The green line indicates a simulation in which we decrease the dispersions by setting $\sigma=0.08$ ($e_{\rm rms}=i_{\rm rms}=0.11$); 2 out 500 become USP planets, while the orange line shows a set where we decrease the masses of the planets by a factor of 3 ($M_{\rm outer}\sim5M_\oplus$). The integrations include GR precession and non-dissipative tidal bulges. }
\label{fig:f_usp}
\end{figure}

\subsection{Formation rate of USP planets}
\label{sec:rate_USP}

We run 500 experiments similar to those in Figure \ref{fig:pop_emax}, but for longer timescales of $30$ Myr and drawing the eccentricities and inclinations of {\it all} planets from a Rayleigh distribution with parameter $\sigma=0.1$ ($e_{\rm rms}=i_{\rm rms}=0.14$). We set the integration time-step to 1 day\footnote{This time-step is somewhat large and barely resolves the pericenter passages when the planet becomes an USP planet ($a_1[1-e_1^2]\simeq 0.02$ AU). However, we checked that by decreasing the time-step to 0.5 days for up to 10 Myr we get consistent results for the fraction of USP planets. We also checked that the secular code Rings, downloadable at
\url{https://github.com/farr/Rings}, gives consistent results.} and stop the simulation when $a_1(1-e_1^2)<0.02$ AU (happening typically at $e_1\sim0.8-0.9$), at which time we assume the planet is tidally captured (Eq. [\ref{eq:a_fmode}]). This is our {\it fiducial} set of integrations. Our goal is to get an estimate of the fraction of systems that can become USP planets and how it depends on the evolution timescale. 

In Figure \ref{fig:f_usp} we show the fraction of USP planets as a function of time (solid blue lines). Recall that we take the condition $a_1(1-e_1^2)<0.02$ AU as a proxy for tidal  capture. We observe that in these integrations the fraction reaches up to $46/500 \simeq9\%$ with most systems reaching the tidal capture threshold after $\sim 1$ Myr ($\sim 100$ to $10^4$ secular cycles, Eq. [\ref{eq:tau_p}]).  The fraction of USP planets does not level off after 30 Myr ($\sim10^9P_1$) and longer integrations are required to assess whether the ensemble reaches a saturation state. We can only say at least $\simeq9\%$ of these Kepler-10-like systems can produce USP planets. We briefly study how this fraction depends on the eccentricity and inclination dispersions as well as the planetary masses.

\paragraph{Lowering eccentricity and inclination dispersions}
We have repeated our fiducial integrations but reduced the eccentricity  and inclination dispersions slightly by changing the parameter from $\sigma=0.1$ ($e_{\rm rms}=i_{\rm rms}=0.11$) to $\sigma=0.08$ ($e_{\rm rms}=i_{\rm rms}=0.11$; green line) and  observe that the final fraction of USP planets decreases from $\simeq9\%$ to $2/500=0.4\%$. We have checked that the systems that undergo fast eccentricity diffusion and become USP planet within 30 Myr are those that have the largest initial values of AMD (or $\sum \left[ e_i^2+i_i^2\right]$; Eq. [\ref{eq:AMD}]). Thus, by slightly reducing  $\sigma$ by $20\%$ we reduce the expected AMD by the same factor. In order to asses whether smaller values of $\sigma$ allow for significant formation of USP planets we need to integrate these systems for Gyr timescales ($10^{11}P_1$). This study is beyond the scope of our paper and it should be the topic of a separate work.

\paragraph{Lowering planetary masses}
We have repeated our fiducial integrations but decreased the masses of all the  planets by a factor of $3$ (orange lines),  so the masses might be more representative of the overall Kepler sample with $M_{\rm USP}\sim1M_\oplus$ and $M_{\rm outer}\sim5M_\oplus$. The fraction of USP planets decreases from $\simeq 9\%$ in the fiducial simulation to  $20/500 \simeq 4\%$. This decrease is expected because the planetary mass scale affects the timescale of the secular evolution ($\tau_{\rm sec}\propto 1/M_{\rm p}$). Thus, the fraction of USP planets in the runs with 3 times lower masses up to 30 Myr of $\simeq 4\%$ should be compared to the fraction in the fiducial simulation up to 10 Myr (i.e., fixed $t_{\rm max}/\tau_{\rm sec}$), which corresponds to  $\simeq 6\%$. The small difference (not statistically significant) between these fractions seems to be due to relativistic precession since it suppresses the diffusion to large eccentricities more efficiently in systems with lower planetary masses.

\begin{figure}[h!]
\center
\includegraphics[width=8.5cm]{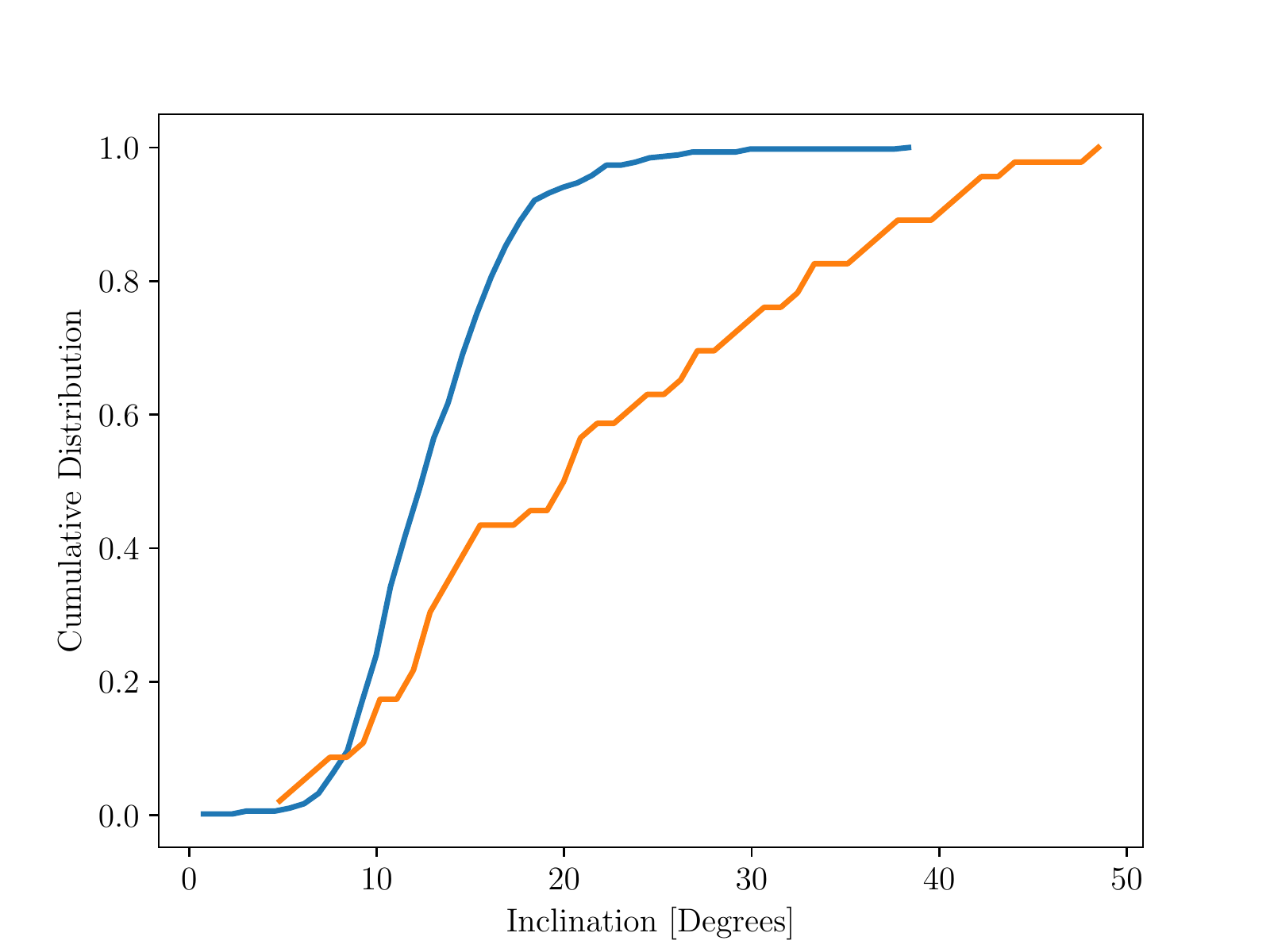}
\caption{Inclination distributions for Kepler-10b from our fiducial simulation (initial $e,i$ from Rayleigh with $\sigma=0.1$, blue line in Figure \ref{fig:f_usp}). The blue line indicates the distribution of the time-averaged inclination for the systems that do not form USP planets, while the orange line indicates the inclinations of systems when they first reach $a_1(1-e_1^2)<0.02$ AU and would be tidally captured.}	
\label{fig:inc_usp}
\end{figure}

\subsection{Inclinations of USP planets}

In Figure \ref{fig:inc_usp} we show the inclinations of the planets that can become USP planets at the moment they first reach $a_1(1-e_1^2)<0.02$ AU and would get tidally capture in our fiducial integrations (orange lines). These inclinations extend from $\sim 5^\circ$ to $\sim 50^\circ$, and in one third of the cases these reach above $\sim 30^\circ$ and are substantially higher than the initial distribution with a median of $\simeq6.7^\circ$ (Rayleigh with $\sigma=0.1$).

\subsection{Non-migrating planets: eccentricities and inclinations}

In Figure \ref{fig:inc_usp} we compute the time-average inclination of the systems that do not form USP planets (blue lines). These systems have an initial distribution with a median of $\simeq6.7^\circ$ (Rayleigh with $\sigma=0.1$) that broadens and reaches a median of $\simeq12^\circ$ as a result of the secular excitation. This implies that the secular gravitational interactions broadens the inclinations of the innermost planets significantly.

\begin{figure}
\center
\includegraphics[width=8.5cm]{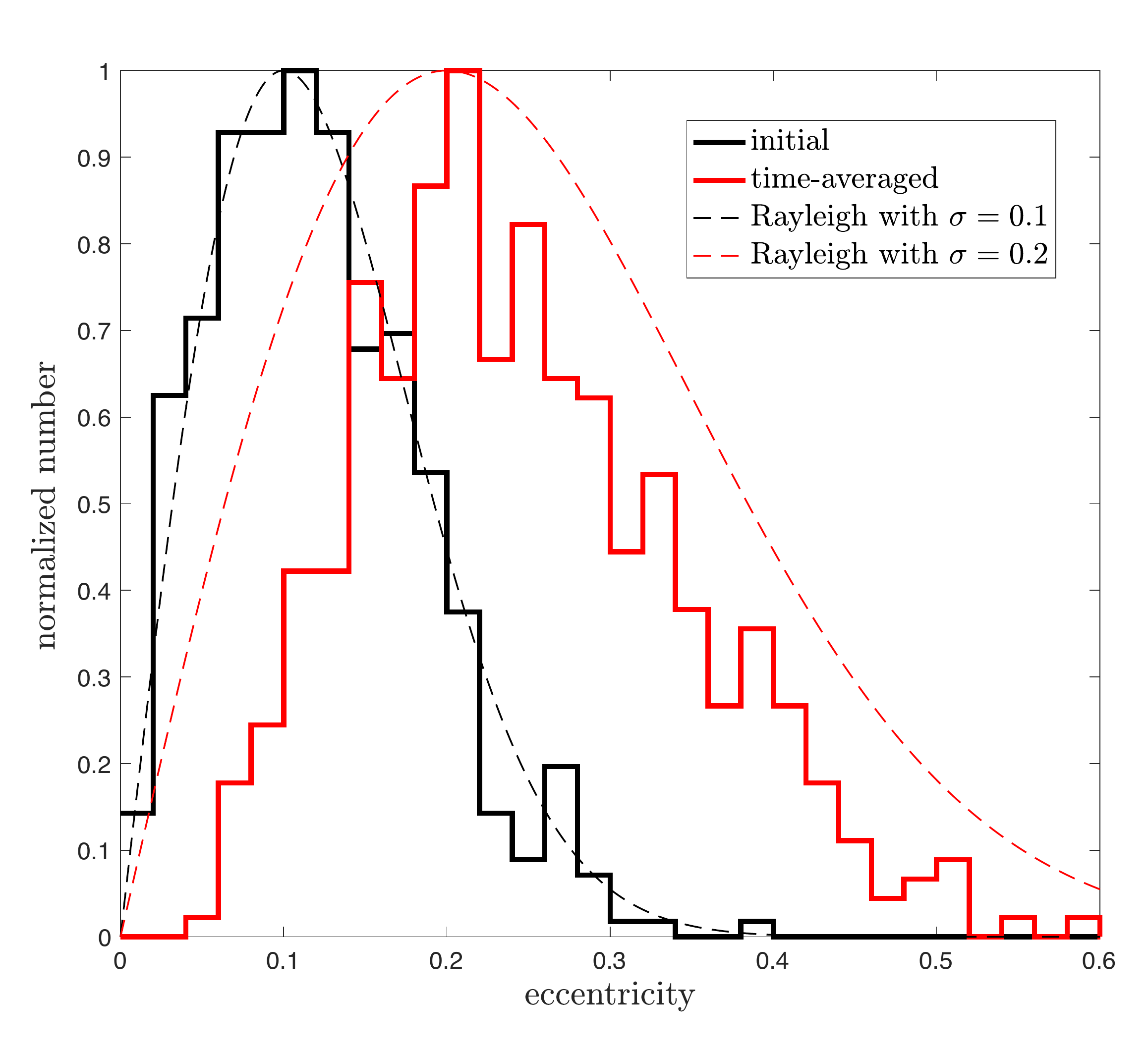}
\caption{Eccentricity distributions for the Kepler-10b planets that do not reach the threshold $a_1(1-e_1^2)<0.02$ AU to become USP planets in our fiducial integrations ($e,i$ from Rayleigh with $\sigma=0.1$ and $t_{\rm max}=30$ Myr). The red line indicates the time-averaged distribution of the ensemble, while the black line shows the initial distribution. For reference the thin dashed lines show the Rayleigh distributions with $\sigma=0.1$ (black) and 
$\sigma=0.2$ (red). The distributions are normalized by the tallest bin.}	
\label{fig:ecc}
\end{figure}

Similarly, in Figure \ref{fig:ecc} we show the time-averaged eccentricity (red line) for the systems without USP planets and compare this with the initial distribution (black line, Rayleigh with $\sigma=0.1$). As it happens with the inclinations, the eccentricities broaden significantly from an initial median of $\simeq 0.11$ to a time-averaged median of $\simeq 0.22$ (the red dashed line indicates a Rayleigh distribution with $\sigma=0.2$ for reference). 

The results above are consistent with the idea that secular chaos can drive the system toward  equipartition of energy of the different secular degrees of freedom (AMD equipartition), where $M_i \sqrt{a_i}e_i^2$ and $M_i \sqrt{a_i}i_i^2$ reach similar values for all $i$ \citep{WL11}. Since the innermost planet has the lowest circular angular momentum (lowest $M_i \sqrt{a_i}$), it gets a larger chunk of the system's eccentricity and inclination budget. In particular, since $M_{\rm outer}/M_{\rm USP}\sim4$ for our example based on Kepler-10, we expect a factor of $\sim2$ increase in the time-averaged eccentricities and inclinations, consistent with the results above.

In summary, we find that secular gravitational interactions leads to excitation of the inner planet's eccentricities and inclinations from an initial  Rayleigh distribution with $\sigma=0.1$ to a time-averaged distribution with $\sigma\simeq0.2$. This result is consistent with secular chaos driving the system towards equipartition of Angular Momentum Deficit.


\section{Discussion} 
\label{sec:discussion}

We propose that most of the ultra-short-period planets around FGK stars are migrated inward by the combined effects of secular chaos and tidal dissipation in the planets. These planets commence their migration from orbital periods beyond $\sim 5$ days.

The key ingredients for our proposal are:
the presence of several planets in the system and a moderate amount of eccentricities and/or inclinations ($e_{\rm rms }, i_{\rm rms}\sim0.1$) to drive chaotic diffusion. As discussed in the introduction, this set of requirements agrees with the observed orbital architecture of planetary systems in the \textit{Kepler} sample. 

In what follows, we discuss various predictions from our model and comment on previous works on this subject.

\subsection{Properties of USP planets from secular chaos}

\subsubsection{Occurrence rate of USP planets}

The observed occurrence rate of USP planets around GK dwarfs is $\sim0.5-1\%$ \citep{SO14}. We consider whether our proposal can explain this rate.  

The occurrence rate of {\it Kepler} systems among the same stars is $\sim 30\%$ \citep{Zhu2018}.  A fraction of these are compact, high-multiple systems that are too tight to be secularly interacting, and are instead dominated by mean-motion resonances. The likely progenitors for USP planets are widely-spaced systems that only contain three or fewer planets within $400$ days. This latter is about half of the overall population \citep{Zhu2018}. For these systems, secular chaos has to produce USP planets at an efficiency of $\sim 5\%$ to account for most of the observed USP planets.

For our fiducial planet architecture, with $e_{\rm rms }, i_{\rm rms}\sim0.1$, and 4 Neptune-massed planets outside the USP planet progenitor,  we find that $10\%$ of the systems have enough AMD to both cause secular chaos and to raise the inner planet's eccentricity toward tidal capture, within our integration time of $30$ Myrs. Due to the diffusive nature of secular chaos, it is expected that this fraction will grow with time, but it is hard to project where it will end up at a few Gyrs.\footnote{If the linear growth (with logarithmic time) as seen in Fig. \ref{fig:f_usp} continues, the final ratio will be $\sim 40\%$.}  In the meantime, our experiments show that dropping the AMD by some $20\%$ (from Rayleigh with $\sigma=0.1$ to 0.08) sharply reduces the yield within 30 Myrs by a factor of $25$, while reducing the masses for the outer companions does not sharply trim down the yield. These considerations argue that, in order for the sparsely spaced {\it Kepler} systems to produce the desired rate of USP planets, the values of eccentricity and inclination dispersion are of order $0.1$, our fiducial value.

Such a dispersion, interestingly, coincides with what is currently determined. \citet{Zhu2018} estimated that sparse systems (with three planets or fewer within $400$ days) have $i_{\rm rms}\sim 0.05-0.1$; and \citep{xie16} argued that $e_{\rm rms}\gtrsim0.1$ for these same systems, which typically appear as single-transiting planets in the {\it Kepler}  database.

So we conclude that, as long as most {\it Kepler} systems contain a fair number of outer companions \citep[as suggested by results from][]{foreman2016,suzuki2016}, our mechanism is likely to account for the observed rate of USP planets.

\subsubsection{Orbital periods: why the 1-day limit for USP planets?}
\label{subsubsec:why1day}

The conventional definition of an USP planet as one that orbits with an orbital period $<1$ day is, in principle, arbitrary. Here we adopt this definition by arguing that there is something physical about the 1-day cut.

In our model, an USP planet is defined as a planet that gets tidally captured. The efficacy of tidal capture drops off steeply beyond a few Roche radii. For instance, tidal capture by $f$-mode excitation leads to the formation of a planet with a final period of $\lesssim 1(\rho_{\oplus}/\rho)^{1/3} \mbox{ day}$, which occurs around $1$ day for planets of Earth density. So we expect planets inward of $1$ day have been placed there by tidal capture, while planets outward should not have experienced this process.  This naturally explain why {\it only} the USP planets are dynamically detached from the companions and why planets at $\sim1-3$ day orbits are less so.

All this being said, it might also be possible to migrate planets to these latter distances ($1-3$ days) via secular chaos. In the case that the innermost planet does not reach the distance for tidal capture but has acquired some substantial eccentricity, tidal circulation is sufficiently efficient that its orbit will decay gradually and it is eventually freed from the influences of other planets.
In fact, Figure \ref{fig:period_ratio} shows a handful of systems with periods in $\sim1-3$ days that have distant companions (5 systems have period ratios of $\gtrsim10$) and for which high-eccentricity might have operated. This possibility should be addressed with a full population synthesis including tidal dissipation.

\subsubsection{Stellar obliquities}

Secular chaos leads to non-linear mixing between the eccentricity and inclination modes, resulting in large excursions in eccentricities and in inclinations.

In our experiments in Figure \ref{fig:inc_usp} we find that the inclinations of the USP planets relative to the initial reference plane can often reach $\sim20^\circ-50^\circ$, much larger than the initial inclination dispersion.

The inclinations reached by USP planets in these experiments are generally lower than previous experiments of secular chaos in hot Jupiter systems by \citet{hamers2017} where planets reach a broad range in $0-140^\circ$ with $\sim10\%$ being retrograde \citep[similar results were found by][]{LW2014}.  We believe that the main difference between the hot Jupiter set-up and ours is that the USP planets start migration much closer in, so these only need to reach $e_1\sim0.9$ in order to migrate. On the other hand, hot Jupiters need to reach $e_1\gtrsim0.98$ and a larger maximum eccentricity might translate into larger attainable inclinations.
 
\subsection{Other Formation Proposals}

We have provided comments on some proposed scenarios for USP planet formation in the introduction section. Here, we provide more detailed assessments on a couple scenario that are similar to ours and involve both dynamical perturbations by other planetary bodies and/or tidal evolution.

\begin{figure}
\center
\includegraphics[width=9.3cm]{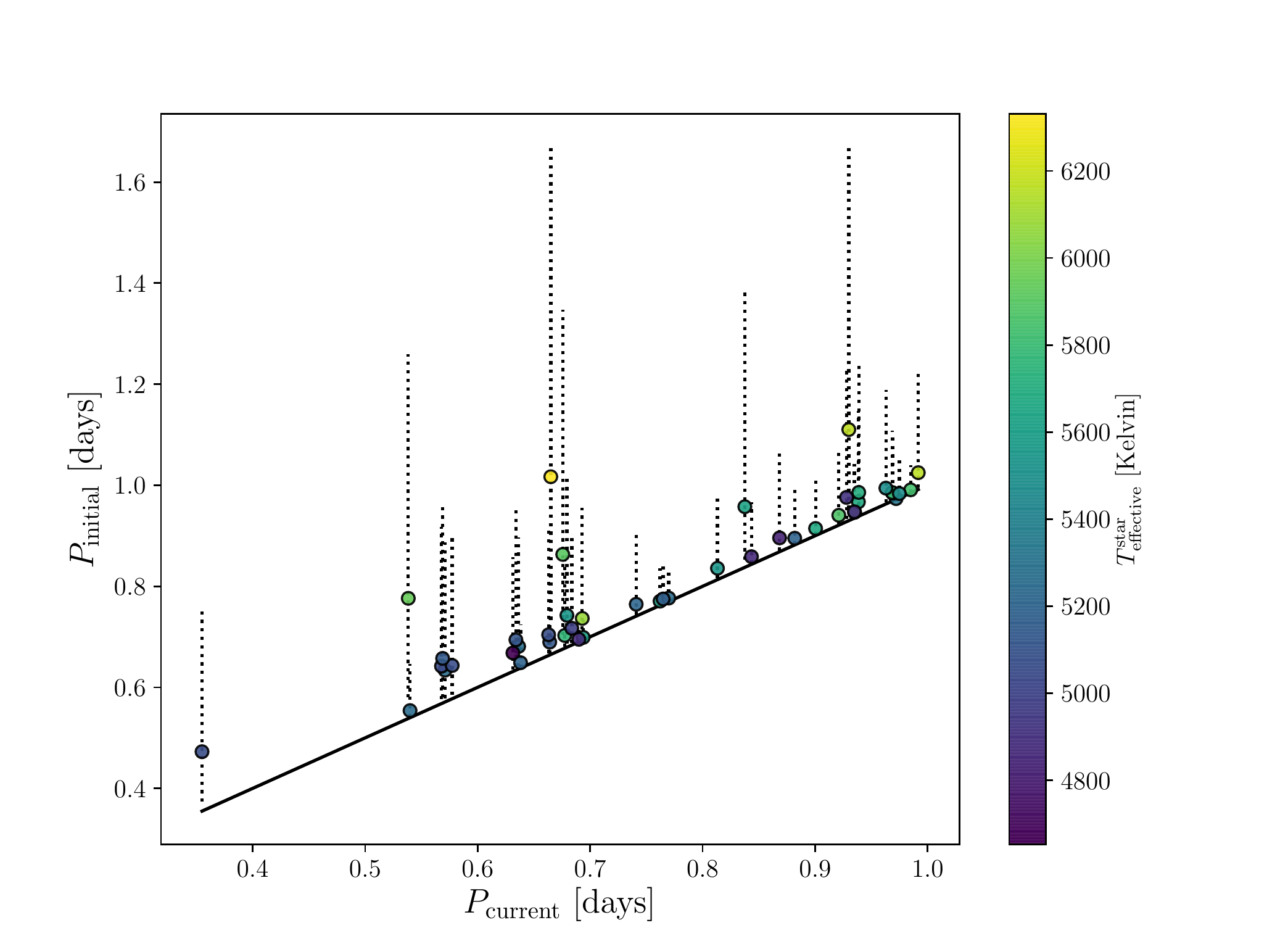}
\caption{Initial period from which the  USP planets started tidal migration if only asynchronous equilibrium tides raised on the star were operating and $P_{\rm spin}\gg 1$ day.  We adopt $Q_s/k_{L,s}=10^7$ and the lines indicate the range $Q_s/k_{L,s}=[10^6,10^8]$, while the lifetime of each system is assumed to be 5 Gyr. The color coding indicates the effective temperature of the host stars.}	
\label{fig:Pin_Pcurrent}
\end{figure}

\subsubsection{Detaching ultra-short-period planets by tides raised on the star}
\label{sec:tide_star}

It was recently suggested by \citet{Lee2017} that the orbits of close-in planets can be eroded from the inner edges of  proto-planetary disks to ultra-short-periods by asynchronous tides raised on their slowly-spinning host stars. This could, in principle, also explain why USP planets are more widely spaced than their longer-period counterparts. We asses this proposal here.

The timescale to shrink its orbit by asynchronous equilibrium tides acting on the host star is (e.g., \citealt{ML02})
\ba
\tau_a^{\rm star}&\equiv&\left|\frac{a}{\dot{a}}\right|=
\frac{P_1}{9\pi}\frac{M_s}{M_1}\left(\frac{a_1}{R_s}\right)^5\frac{Q_s}{k_{L,s}}
\left|1-\frac{P_1}{P_{\rm spin}}\right|^{-1}
\label{eq:tau_s_star}
\ea
where $Q_s$ and $k_{L,s}$ are tidal quality factor and the Love number of the star. For reference, for a Sun-like star with $P_{\rm spin}\gg1$ day one obtains
\ba
\tau_a^{\rm star}
&\simeq&5\times10^{10}\mbox{ yr }\left(\frac{P_1}{1 \mbox{day}}\right)^{13/3}
\left(\frac{1M_\oplus}{M_1}\right) \left(\frac{1R_\odot}{R_s}\right)^5\nonumber\\
&&\times\left( \frac{M_s}{1M_\odot} \right)\left(\frac{Q_s/k_{L,s}}{10^6}\right),
\label{eq:tau_star_2}
\ea
which is an order of magnitude longer than the typical ages of the host stars and typical values of the modified quality factor constrained from observations  $Q_s/k_{L,s}\gtrsim10^6$ \citep[e.g.,][]{ogilvie2014}

We further  quantify the extent at which tides on the star can shrink the orbits of USP planets. We integrate Equation (\ref{eq:tau_s_star}) backwards in time and estimate the initial period $P_{\rm initial}$ the planets would have had in order to reach their current locations $P_{\rm current}$ after a time $t_{\rm age}$:
\ba
P_{\rm 1, initial}=P_{\rm 1, current}\left(1+\frac{13}{2}\frac{t_{\rm age}}{\tau_{\rm a, current}^{\rm star}},\right)^{2/13}.
\ea
where $\tau_{\rm a, current}^{\rm star}$ is given by Equation (\ref{eq:tau_star_2}) at their current location.

In Figure \ref{fig:Pin_Pcurrent}  we show the initial period $P_{\rm initial}$ for our sample of USP planets assuming $Q_s/k_{L,s}=10^7$ (circles) and the dashed lines indicate the range $Q_s/k_{L,s}=[10^6,10^8]$.  We adopt $t_{\rm age}=5$Gyr and for the systems in which there is not a mass estimate (all but two systems), we use the mass-radius relation for cold, terrestrial-mass planets from \citealt{Seager07}.  The values of of $Q_s/k_{L,s}$ cover the wide range that has been inferred through ensemble  analyses of star-planet systems (e.g., \citealt{hansen10,penev12,CJ18}). In particular, the recent work by \citet{penev18} looking at the tidal spin up of Sun-like stars with hot Jupiters constrains $Q_s/k_{L,s}\sim10^7$ for orbital periods of $\sim1$ day.

We observe that only $3/44\sim7\%$ ($15/44\sim34\%$) of the systems could have migrated from $\gtrsim1$ day to their current location inside 1 day for $Q_s/k_{L,s}=10^7$ ($Q_s/k_{L,s}=10^6$).  Even with $Q_s/k_{L,s}=10^6$, the changes in periods are generally modest and the handful of planet with the most dramatic changes in periods tend to have higher effective temperatures, possibly indicating larger $Q_s/k_{L,s}$ since hotter stars have thinner convective envelopes \citep[e.g.,][]{ogilvie2014}.

We conclude that equilibrium tides on the star are unlikely to shrink the planetary orbits from initial periods of $\gtrsim1$ days to their current locations and are, therefore, unable to account for the observed detachment from their outer companions. However, as pointed out by  \citet{Lee2017}, tides on the star can do a good job at reproducing the steep fall\footnote{In steady-state the equation (\ref{eq:tau_star_2}) predicts that $dN/dP\propto P^{10/3}$.} in frequency of USP planets with decreasing period observed by \citet{SO14}. It is possible that the data are best explained by a combination of secular chaos delivering planets to $\sim1$ day orbits and subsequent sculpting of the period distribution from tides raised on the star.

\subsubsection{Low-eccentricity migration in muti-planet systems}

Another possibility is that USP planets may have migrated inwards due to gravitational interactions with outer companions and gradual dissipation of orbital energy by tides acting on the planets  \citep{ML04,Schlaufman2010}. These systems follow low-eccentricity paths during migration, different from our proposed model invoking high eccentricities.

For small eccentricities $\mathcal{F}(e)\simeq7e^2/2$ in Equation (\ref{eq:F_e}) and $a_{\rm f}\simeq a$. Thus, Equation (\ref{eq:tau_a_hut}) becomes:
\ba
\tau_{a}^{\rm planet}&\simeq&1.3\times10^9~\mbox{yr}~\left(\frac{Q_p/k_{L,p}}{100}\right)
\left(\frac{1R_\oplus}{R_1}\right)^5
\left(\frac{M_1}{1M_\oplus}\right)\nonumber\\
&&\times\left(\frac{P_1}{1~\mbox{day}}\right)^{13/3}
\left(\frac{0.01}{e_1}\right)^{2}.
\label{eq:tau_a_lowe}
\ea
This expression implies that the planet moves inwards as long as it can maintain an eccentricity at the percent level, which can be forced secularly\footnote{Mean-motion resonances can also force eccentricities against tidal dissipation, but the outer planets would have periods of less than a few days in this scenario, being unable to match the observations (Fig. \ref{fig:g_out_kappa}).}  by outer planets.

In this picture, the inner planet is subject to tidal dissipation, secular forcing from other planets, and apsidal precession from GR (and possibly other sources of precession). For a two-planet system \citet{ML04} argued that the evolution behaves like that of a damped autonomous system that tends to align their apsidal orientations and reach a quasi-fixed point. At this state, the inner planet reaches an equilibrium eccentricity given by \citep{M07}:
\ba
e_1^{\rm eq}=\frac{5}{4}\frac{(a_1/a_2)e_{\rm 2}}{|1-(M_1/M_2)(a_1/a_2)^{1/2}+\gamma|},
\ea
where 
\ba
\gamma&\equiv&\frac{4GM_s}{c^2a}\frac{M_s}{M_{\rm 2}}\left(\frac{a_2}{a_1}\right)^3\frac{1}{(1-e_1^2)}\nonumber\\
&\simeq&5.6\left(\frac{10M_\oplus}{M_2}\right)
\left(\frac{1\;\mbox{day}}{P_1}\right)^{8/3}
\left(\frac{P_2}{10\;\mbox{day}}\right)^{2}.
\label{eq:gamma}
\ea
The fractional contribution that the relativistic potential makes to the apsidal advance, compared to the outer planet, is $\gamma/(1+\gamma)$. Thus, for a given perturber at $a_2$ with eccentricity $e_2$ we estimate the equilibrium eccentricity and replace  this into $\tau_{a}^{\rm planet}$ in Equation (\ref{eq:tau_a_lowe}) to estimate  the migration timescale.

For $\gamma\gg1$, we can approximate $e^{\rm eq}\propto 1/\gamma$, so we can plug $e^{\rm eq}$ into Equation (\ref{eq:tau_a_lowe}) to obtain 
\ba
&&\tau_{a,{\rm eq}}^{\rm planet}\sim6\times10^9~\mbox{yr}~\left(\frac{Q_p/k_{L,p}}{100}\right)
\left(\frac{1R_\oplus}{R_1}\right)^5
\left(\frac{M_1}{1M_\oplus}\right)\nonumber\\
&&\times\left(\frac{P_1}{1~\mbox{day}}\right)^{-7/3}\left(\frac{P_2}{10\;\mbox{day}}\right)^{16/3}
\left(\frac{10M_\oplus}{M_2}\right)^2\left(\frac{0.1}{e_2}\right)^{2}.
\label{eq:tau_a_lowe_2}
\ea
From this expression we conclude that this migration channel can work for $M_2\lesssim10M_\oplus$ only if $P_2\lesssim10$ days. One example of a planetary system in this regime is CoRoT-7  which has two super-Earth planets inside 10 days  \citep{leger2009}. For Jupiter-mass planets the model  can work for companions in much wider orbits ($P_2\lesssim35$ days). Consistent with these estimates, \citet{HM15} found that linear secular forcing in multi-planet systems is unable produce USP planets for $P_2\gtrsim 10$ days.

Based on the observed multiplicity in the \textit{Kepler} sample, we argued in  \S\ref{sec:data} that most USP planets have {\it Kepler}-like companions beyond $\sim20$ days.  The possibility of having giants is disfavored by the lack of a metallicity trend for the USP planets' host stars \citep{Winn2017}. We conclude that low-eccentricity migration can account for some USP planets in relatively more compact configurations, but possibly not the majority.

\subsection{Predictions for {\it TESS}}

The soon-to-be launched Transiting Exoplanet Survey Satellite (\textit{TESS}) by NASA is expected to find a couple dozen USP planets and hundreds of small planets ($\lesssim 2R_\oplus$) in short-period orbits ($\lesssim10$ days) around FGK stars \citep{sullivan2015}. While a smaller sample size than that of {\it Kepler}, the {\it TESS} sample has the advantage that the host stars are much brighter and closer-in.  This sample might allow testing some of our model predictions.

\begin{itemize} 

\item For fast rotating host stars, the {\it spin-orbit angle} can potentially be measured using the Rossiter-McLaughlin effect, as has been attempted for the USP $\rho$ 55 Cancri e \citep{BH2014,LM2014}. Similarly, measurements of the orientation of the star's rotation axis can constrain the obliquity angles. Various techniques have been discussed, including ensemble analysis using projected rotational velocities \citep{Schlaufman2010b,MW2014} and individual systems using asteroseismology \citep{huber2013} or stellar  spot crossing \citep{deming11,SW11}.  A systematic misalignment (e.g., misalignment angle above $\sim20^\circ$) would strongly support our model, especially if their longer-period counterparts have small misalignment angles.

\item Follow-up radial velocity campaigns of USP planet systems  may detect non-transiting companions in tens of days orbits, which are predicted by our model. We expect these {\it companions} to be typically more massive than the USP planets.

\item Radial velocity follow-up of low-mass \textit{TESS} planets (non-USP planets) may also determine their orbital eccentricities, a key ingredient in our model. We require that planets in low-multiple systems tend to have larger eccentricities, with the innermost planet having diffused to even higher eccentricities due to secular interactions ($e\sim0.2-0.4$). These planets should orbit beyond $P \sim10$ days so tides do not generally damp their eccentricities ($\tau_a^{\rm planet}\gg10$ Gyr in Eq. [\ref{eq:tau_a_lowe}]). We also expect to observe a population of {\it eccentric planets}  in short-period orbits ($\lesssim10$ days)  that are excited by secular chaos with neighbouring planets and might have suffered some tidal decay as a consequence, but have yet to attain so high an eccentricity to experience tidal capture (\S \ref{subsubsec:why1day}).

\end{itemize}


\section{Summary} \label{sec:conclusion}

If a large number of {\it Kepler} planetary systems contain a fair number of planets, possibly extending to $\gtrsim \mbox{AU}$ distances, and their orbits have  non-negligible eccentricities and inclinations, we argue here that these systems can undergo large eccentricity and inclinations variations by non-linear secular interactions, through a process termed secular chaos. One of the consequences is the production of an ultra-short-period (USP) planet, as when the innermost planet (from periods of $5-10$ days) acquires high eccentricities from the interactions, it can approach the host star at such a close range that it is tidally captured onto a very short orbit (period short-ward of a day).
We propose that most USP planets are formed by tidal high-eccentricity migration driven by secular chaos from periods of $\sim5-10$ days.

Such a mechanism is intimately related to one of the mechanisms that has been proposed for forming hot Jupiters. In fact, the USP planets can be thought of as a scaled-down version of the hot Jupiter formation, both in planet masses, and in distance scales. For the latter, USP planets start their migration from around $\lesssim0.1$ AU, while hot Jupiters from periods outward of $\sim 1$ AU.  So while the latter need to reach extreme eccentricities ($e \gtrsim 0.98$) to be tidally captured, a lower value ($e\gtrsim 0.8$) is required for making USP planets. 

Our scenario naturally explains the observation that most USP planets have distant companions with periods of $\sim 10- 50$ days (Figs. 2 and 3), while their  (not ultra-)short-period counterparts ($\sim1-3$ days) reside in more compact configurations. We predict that USP planets orbit in inclined ($\sim20^\circ-50^\circ$) planes relative to those of both their outer companions and their host star's equator.

This scenario, even in cases where the eccentricity acquired by the innermost planet is  not sufficiently high to be tidally captured, should, through gradual tidal circularization, bring the planets closer to the stars, perhaps forming some of the short-period planets (period of a couple days). 

Future discoveries from {\it TESS} along with follow-up studies of systems with USP planets will reveal the prevalence of secular chaos in shaping the dynamical properties  of USP planets.

Interestingly, the production of an USP planet could also happen in our very own Solar System in the future.  The orbital evolution of Mercury is known to be chaotic with diffusion of the eccentricities on  billion-year timescales. This diffusion can lead to Mercury colliding with Venus or the Sun within the next five billion years, before the Sun becomes a red giant \citep[at $1\%$ probability,][]{LG2009}. Except in the latter case Mercury will not collide with the Sun -- as its eccentricity gradually rises due to secular interactions, tidal capture will snare it away from the influences of other planets. A new ultra-short-period planet will be born and will become detectable by {\it Kepler} missions launched by alien civilizations.

\section*{Acknowledgements}
We are grateful to Daniel Tamayo, Diego Mu{\~n}oz, Eve Lee, Norm Murray, and Wei Zhu for enlightening discussions.  C.P. acknowledges support from the Gruber Foundation Fellowship and Jeffrey L. Bishop Fellowship. Y. W. thanks NSERC for research support.  This paper includes data collected by the Kepler mission. Funding for the Kepler mission is provided by the NASA Science Mission directorate.  This research has made use of the NASA Exoplanet Archive, which is operated by the California Institute of Technology, under contract with the National Aeronautics and Space Administration under the Exoplanet Exploration Program.

\appendix
\section{Transit probabilities in multi-planet systems from \citet{TD11}}

We summarize the method by \citet{TD11}
to estimate the transit probabilities in multi-planet systems.
We only provide with the necessary details to reproduce our results
in \S\ref{sec:data} and invite the reader to consult \citet{TD11} for full explanations as we 
limit ourselves to providing the minimal amount of details required to reproduce 
our results.

As in  \citet{TD11} we consider a system containing $n$ planets with semi-major axes 
specified by $\epsilon_{i}=R_s/a_i$ with $i=1,..,n$.
 The probability that a randomly oriented observer will detect $m$ transiting planets 
 in this system is defined by $g_{\rm mn}(\epsilon_{1},...,\epsilon_{n})$.
 For a single planet system we have the usual expression
 \ba
 g_{11}(\epsilon)=\epsilon=\frac{R_s}{a}.
 \ea
 For $n>1$ we 
 assume that the inclinations $i$ relative to an arbitrary reference plane (e.g., the invariable
 plane) is given by a Fischer distribution
\begin{equation} \label{eqn:fischer}
    q(i|\kappa) = \frac{\kappa \sin{i}}{2\sinh{\kappa}} e^{\kappa\cos{i}}\ , 
\end{equation}
where the parameter $\kappa$ is related to the mean-square value of $\sin{i}$ 
\begin{equation} \label{eqn:kappa}
    \langle \sin^2{i} \rangle = \frac{2}{\kappa} \left(\coth{\kappa}-\frac{1}{\kappa}\right)\ .
\end{equation}
 
 Then, the probability of a transit of a single planet, given
the observer orientation $\cos \theta$
 is
\begin{align}
u(x|\epsilon,\kappa)= \sum_{\ell=0}^\infty Q_\ell(\kappa)\,b_\ell(\epsilon) P_\ell(\cos \theta).
\label{eq:one}
\end{align}
where 
$P_\ell$ is a Legendre polynomial,
\ba
Q_\ell(\kappa)&\equiv& \int_0^\pi di\,q(i|\kappa)P_\ell(\cos i), \quad Q_0=1,\nonumber\\
&=&\sqrt\frac{\pi\kappa}{2}\frac{I_{\ell+1/2}(\kappa)}{\sinh\kappa}
\ea
with $I$ denoting a modified Bessel function
and
\begin{equation}
b_\ell(\epsilon)=\left\{\begin{array}{ll}\epsilon, & \ell=0, \\
P_{\ell+1}(\epsilon)-P_{\ell-1}(\epsilon), &
  \ell\hbox{ even, } \ell>0 \\
              0, & \ell\hbox{ odd.}\end{array}\right.
\end{equation}

\paragraph{Two-planet system} Assuming that the intrinsic
multiplicity is $n=2$, 
the probability that both transit
for a random orientation of the observer is 
\ba
\label{eq:g22}
g_{22}(\epsilon_1,\epsilon_2,\kappa)&=& \frac{1}{2}\int_{-1}^1 d\cos \theta\, u(\cos \theta|\epsilon_1,\kappa)u(\cos \theta|\epsilon_2,\kappa) \nonumber\\
&=&\sum_{\ell=0}^\infty \frac{Q_\ell^2(\kappa)}{2\ell+1}b_\ell(\epsilon_1)b_\ell(\epsilon_2).
\ea
By denoting the inner planet as planet 1, $\epsilon_1>\epsilon_2$, 
the probability that only the inner planet transits is
\ba
g_{12}^{\rm inner}(\epsilon_1,\epsilon_2)&=& \frac{1}{2} \int_{-1}^1 d\cos \theta \, u(\cos \theta|\epsilon_1,\kappa)
\left[1-u(\cos \theta|\epsilon_2,\kappa)\right] \nonumber\\
&=& g_{11}(\epsilon_1)-g_{22}(\epsilon_1,\epsilon_2,\kappa).
\ea

Thus, for an intrinsic two-planet system 
we can calculate the expected ratio of between the number of USP planets
observed as single-transiting and those observed as multi-transiting systems
as
\ba
\left<\frac{f_{>1}}{f_{1}}\right>
=\frac{g_{22}(\epsilon_1,\epsilon_2,\kappa)}{\epsilon_1-
g_{22}(\epsilon_1,\epsilon_2,\kappa)}
\label{eq:g_ups_1}
\ea

 \paragraph{Three or more planet systems}
 The expression in Equation (\ref{eq:g_sup}) can be extended to 
 higher intrinsic multiplicity systems using the formalism 
 by  \citet{TD11}.
 Evidently, higher planet multiplicities
will translate in higher values of $<f_{>1}^{\rm USP}/f_{1}>$.
 We have not performed this analysis as it adds
an extra degree of freedom and it requires knowledge 
of the intrinsic multiplicity of systems with USP planets.



\begin{thebibliography}{}
\bibitem[Adams et al.(2016)]{adams16}
Adams E. R., Jackson B., Endl M., 2016, \apj, 152, 47
\bibitem[Batalha et al.(2011)]{Batalha2011} Batalha, N. M., Borucki, W. J., Bryson, S. T., et al. 2011, \apj, 729, 27
\bibitem[Bourier \& Henrard(2014)]{BH2014} Bourier V., \& Henrard G., 2014, \aap, 569, A65
\bibitem[Burke et al.(2015)]{burke2015}
Burke, C. J., Christiansen, J. L., Mullally, F., et al. 2015, ApJ, 809, 8
\bibitem[Buchhave et al.(2012)]{Buchhave2012} Buchhave, L. A., Latham, D. W., Johansen, A., et al. 2012, Nature, 486, 375
\bibitem[Butler et al.(1997)]{Butler1997} Butler, R. P., Marcy, G. W., Williams, E., et al. 1997, \apj, 474, 115
\bibitem[Collier Cameron \& Jardine(2018)]{CJ18}
Collier Cameron, Andrew \& Jardine, Moira 2018, \mnras, arXiv:1801.10561 
\bibitem[Chiang \& Laughlin(2013)]{Chiang2013} 
Chiang, E., \& Laughlin, G. 2013, \mnras, 431, 3444
\bibitem[D'Antona \& Mazzitelli(1994)]{DAntona1994} D'Antona, F., \& Mazzitelli, I. 1994, \apjs, 90, 467
\bibitem[Dawson \&  Johnson(2018)]{DJ2018}
Dawson, R. I., \&  Johnson, J. A. 2018, arXiv:1801.06117
\bibitem[Deming et al.(2011)]{deming11}
Deming, D., P. V. Sada, B. Jackson, S. W.et al. 2011, ApJ, 740, 33
\bibitem[Dressing \& Charbonneau(2015)]{Dressing} Dressing, C.~D., \& Charbonneau, D.\ 2015, \apj, 807, 45
\bibitem[Dumusque et al.(2014)]{dum2014}
Dumusque, X., Bonomo, A. S., Haywood, R. D., et al. 2014,
\bibitem[Fischer \& Valenti(2005)]{Fischer2005} Fischer, D. A., \& Valenti, J. 2005, \apj, 622, 1102
\bibitem[Fischer et al.(2008)]{Fischer2008} Fischer, D. A., Marcy, G. W., Butler, R. P., et al. 2008, \apj, 675, 790
\bibitem[Foreman-Mackey et al.(2016)]{foreman2016} Foreman-Mackey D., Morton T. D., Hogg D. W., et al.
2016, AJ, 152, 206
\bibitem[Fulton et al.(2017)]{fulton2017}
Fulton, B. J., Petigura, E. A., Howard, A. W., et al. 2017, AJ, 154, 109
\bibitem[Gonzalez(1997)]{Gonzalez1997} Gonzalez, G., 1997, \mnras, 285, 403
\bibitem[Hamers et al.(2017)]{hamers2017}
Hamers, A. S., Antonini, F., Lithwick, Y., Perets, H. B., \& Portegies Zwart, S. F.
2017, MNRAS, 464, 688
\bibitem[Hansen(2010)]{hansen10}
Hansen, B. M. S. 2010, ApJ, 723, 285
\bibitem[Hansen \& Murray(2015)]{HM15}
Hansen, B. M. S., \& Murray, N. 2015, MNRAS, 448, 1044
\bibitem[Hansen \& Zink(2015)]{HZ2015}
Hansen, B. M. S., \& Zink, J. 2015, MNRAS, 450, 4505
\bibitem[Howard et al.(2012)]{howard2012}
Howard, A. W., Marcy, G. W., Bryson, S. T., et al. 2012, ApJS, 201, 15
\bibitem[Huang et al.(2016)]{huang2016}
Huang, C., Wu, Y., \& Triaud, A. H. M. J. 2016, ApJ, 825, 98
\bibitem[Huber et al.(2013)]{huber2013}
Huber, D., Carter, J. A., Barbieri, M., et al. 2013, Sci, 342, 331
\bibitem[Hut(1981)]{hut81}
Hut P., 1981, \aap, 99, 126
\bibitem[Jackson et al.(2013)]{Jackson2013} Jackson, B., Stark, C. C., Adams, E. R., et al. 2013, \apj, 779, 165
\bibitem[Jackson et al.(2016)]{Jackson2016} Jackson, B., Jensen, E., Peacock, S., et al. 2016, CeMDA, 126, 227
\bibitem[Knutson et al.(2014)]{Knutson14} Knutson, H. A., Fulton, B. J., Montet, B. T., et al. 2014, \apj, 785, 126
\bibitem[Lambeck(1977)]{lambeck77}
Lambeck, K., 1977, Philosophical Transactions of the Royal Society 
of London Series A, 287, 545
\bibitem[Laskar(1996)]{laskar96} Laskar, J. 1996, CeMDA, 64, 115
\bibitem[Laskar(1997)]{laskar97}
Laskar, J. 1997, A\&A, 317, L75
\bibitem[Laskar \& Gastineau(2009)]{LG2009}
Laskar, J., \& Gastineau, M. 2009, Nature, 459, 817
\bibitem[Lee \& Chiang(2017)]{Lee2017} Lee, E. J., \& Chiang, E. 2017, \apj, 842, 40
\bibitem[L\'eger et al.(2009)]{leger2009}
L\'eger, A., Rouan, D., Schneider, J., et al. 2009, A\&A, 506, 287
\bibitem[Lithwick \& Wu(2011)]{LW2011}
Lithwick, Y., \& Wu, Y. 2011, \apj, 739, 31
\bibitem[Lithwick \& Wu(2014)]{LW2014}
Lithwick, Y., \& Wu, Y. 2014, PNAS, 111, 12610
\bibitem[Liu et al.(2015)]{liu15} 
Liu, B., Mu{\~n}oz, D.~J., \& Lai, D.\ 2015, \mnras, 447, 747 
\bibitem[L\'opez-Morales et al.(2014)]{LM2014}
L\'opez-Morales, M.; Triaud, A. H. M. J., Rodler, F. et al. 
2014, ApJ, 792, L31
\bibitem[Lundksvist et al.(2016)]{Lundkvist2016} Lundkvist, M. S., Kjeldsen, H., Albrecht, S., et al. 2016, Nature Communications, 7, 11201
\bibitem[Mandell et al.(2007)]{Mandell2007} Mandell, A. M., Raymond, S. N., \& Sigurdsson, S. 2007, \apj, 660, 823b
\bibitem[Marcy et al.(2002)]{Marcy2002} Marcy, G. W., Butler, R. P., Fischer, D. A., et al. 2002, \apj, 581, 1375
\bibitem[Marcy et al.(2014)]{marcy2014}
Marcy, G. W., Isaacson, H., Howard, A. W., Rowe, et al. 2014, ApJS, 210, 20
\bibitem[Mardling \& Lin(2002)]{ML02}
Mardling, R., \& Lin, D. N. C. 2002, ApJ, 573, 829
\bibitem[Mardling \& Lin(2004)]{ML04}
Mardling R. A., \& Lin D. N. C., 2004, ApJ, 614, 955
\bibitem[Mardling(2007)]{M07}
Mardling, R., 2007, MNRAS, 382, 1768
\bibitem[Mayor et al.(2011)]{mayor2011}
Mayor, M., Marmier, M., Lovis, C., et al. 2011, arXiv:1109.2497
\bibitem[McArthur et al.(2004)]{McArthur2004} 
McArthur, B. E., Endl, M., Cochran, W. D., et al. 2004, \apj, 614, 81
\bibitem[Morton \& Winn(2014)]{MW2014}
Morton, T. D., \& Winn, J. N. 2014, ApJ, 796, 47
\bibitem[Mu{\~n}oz et al.(2016)]{munoz2016}
Mu{\~n}oz, D. J., Lai, D., \& Liu, B. 2016, MNRAS, 460, 1086
\bibitem[Murray-Clay et al.(2009)]{Murray-Clay09} Murray-Clay, R.~A., Chiang, E.~I., \& Murray, N.\ 2009, \apj, 693, 23 
\bibitem[Nelson et al.(2014)]{nelson14}
Nelson, B. E.,  Ford, E. B.,  Wright, J. T. Fischer, D. A., et al. , MNRAS , 2014, 441, 442
\bibitem[Neron de Surgy \& Laskar(1997)]{neron97}
Neron de Surgy, O.,\&  Laskar, J., 1997, \aap, 318, 975
\bibitem[Ogilvie(2014)]{ogilvie2014}
Ogilvie, G. I. 2014, ARA\&A, 52, 171
\bibitem[Owen \& Wu(2013)]{Owen2013} Owen, J. E., \& Wu, Y. 2013, \apj, 775, 105
\bibitem[Palla \& Stahler(1991)]{Palla1991} Palla, F., \& Stahler, S. W. 1991, \apj, 375, 288
\bibitem[Penev et al.(2012)]{penev12}
Penev, K., Jackson, B., Spada, F., \& Thom, N. 2012, ApJ, 751, 96
\bibitem[Penev et al.(2018)]{penev18}
Penev, K., Bouma, LG., Winn, J. N, \& Hartman, J. D.  2018, arXiv:1802.05269
\bibitem[Petigura et al.(2017)]{petigura2017}Petigura, E. A., Marcy, G. W., Winn, J. N., et al. 2017b, ArXiv
e-prints, arXiv:1712.04042
\bibitem[Puranam \& Batygin(2018)]{PB2018}
Puranam, A. \& Batygin, K. 2018, arXiv:1802.08385
\bibitem[Queloz et al.(2009)]{queloz2009}
Queloz, D., Bouchy, F., Moutou, C., et al. 2009, \aap, 506, 303
\bibitem[Rein \& Liu(2011)]{Rein2011} 
Rein, H., Liu, S-F., 2011, \apj, 527 
\bibitem[Rein \& Tamayo(2015)]{RT15} 
Rein, H., \& Tamayo, D.\ 2015, \mnras, 452, 376 
\bibitem[Rice(2015)]{rice2015}
Rice, K. 2015, MNRAS, 448, 1729
\bibitem[Sanchis-Ojeda et al.(2014)]{SO14} 
Sanchis-Ojeda, R., Rappaport, S., Winn, J. N., et al. 2014, \apj, 787, 47
\bibitem[Sanchis-Ojeda \& Winn(2011)]{SW11} 
Sanchis-Ojeda, R. \& Winn, J. N., \apj, 743, 61
\bibitem[Santos et al.(2004)]{Santos2004} 
Santos, N. C., Israelian, G., \& Mayor, M. 2004, \aap, 415, 1153
\bibitem[Schlaufman et al.(2010)]{Schlaufman2010} 
Schlaufman, K. C., Lin, D. N. C. \& Ida, S. 2010, \apj, 724, L53
\bibitem[Schlaufman(2010)]{Schlaufman2010b} 
Schlaufman, K. C. 2010, ApJ, 719, 602
\bibitem[Schlaufman \& Laughlin(2011)]{Schlaufman2011} 
Schlaufman, K. C., \& Laughlin, G. 2011, \apj, 738, 177
\bibitem[Seager et al.(2007)]{Seager07} 
Seager, S., Kuchner, M., Hier-Majumder, C.A., \& Militzer, B. 2007, \apj, 669, 1279
\bibitem[Steffen et al.(2012)]{Steffen2012} Steffen, J.~H., Ragozzine, D., Fabrycky, D.~C., et al.\ 2012, PNAS, 109, 7982 
\bibitem[Steffen \& Farr(2013)]{SF13}
Steffen, J. H., \& Farr, W. M. 2013, ApJL, 774, L12
\bibitem[Steffen \& Coughlin(2016)]{Steffen2016} 
Steffen, J. H., \& Coughlin, J. L. 2016, PNAS, 113, 43 
\bibitem[Steffen \& Hwang(2015)]{SH15}
Steffen, J. H., \& Hwang, J. A. 2015, MNRAS, 448, 1956
\bibitem[Sterne(1939)]{Sterne39} Sterne, T.~E.\ 1939, \mnras, 99, 451 
\bibitem[Sullivan et al.(2015)]{sullivan2015}
Sullivan, P. W., Winn, J. N., Berta-Thompson, Z. K. 2015, ApJ, 809, 77
\bibitem[Suzuki et al.(2016)]{suzuki2016}
Suzuki, D., Bennett, D.P., Sumi, T., et al. 2016, ApJ, 833, 145
\bibitem[Swift et al.(2013)]{swift2017}
Swift, J., Johnson, J. A., Morton, T. D., Crepp J. R., et al.
 2013, ApJ, 764, 105
\bibitem[Tremaine \& Dong(2011)]{TD11} Tremaine, S., \& Dong, S. 2011, \aj, 143, 94
\bibitem[Udry et al.(2006)]{Udry2006} Udry, S., Mayor, M., Benz, W., et al. 2006, \aap, 447, 361
\bibitem[Valencia et al.(2010)]{Valencia2010} Valencia, D., Ikoma, M., Guillot, T., et al. 2010, \aa, 516, A20
\bibitem[Valsecchi et al.(2014)]{Valsecchi2014} Valsecchi, F., Frederic, A. R., \& Steffen, J. H. 2014, \apjlett, 793, 3
\bibitem[Vick \& Lai(2018)]{VL2018}
Vick, M., \& Lai, D. 2018, MNRAS, 
\bibitem[Weiss et al.(2016)]{weiss2016}
Weiss, L. M., Rogers, L. A., Isaacson, H. T., et al. 2016, ApJ,
819, 83
\bibitem[Winn et al.(2017)]{Winn2017} Winn, J. N., Sanchis-Ojeda, R., Rogers, L., et al. 2017, \aj, 154, 60
\bibitem[Wright et al.(2012)]{Wright12} Wright, J.~T., Marcy, G.~W., Howard, A.~W., et al.\ 2012, \apj, 753, 160 
\bibitem[Wu(2018)]{Wu2018}
Wu, Y. 2018, AJ 155, 118 
\bibitem[Wu \& Lithwick(2011)]{WL11} Wu, Y. \& Lithwick, Y. 2011, \apj, 735,109
\bibitem[Xie et al.(2016)]{xie16} 
Xie, J.-W., Dong, S., Zhu, Z., et al. 2016, PNAS, 113, 11431
\bibitem[Zhu et al.(2018)]{Zhu2018} Zhu, W., Petrovich, C., Wu, Y. et al. 2018, arXiv:1802.09526

\end{thebibliography}
\end{document}